\documentclass[prb,showkeys]{revtex4}
\usepackage{color,graphicx,amsmath,amssymb,bm,epsfig}

\begin{document}

\title{Properties of mesoscopic hybrid superconducting systems}

\author{Fabio Taddei\footnote{E-mail: f.taddei@sns.it}, Francesco Giazotto, and Rosario Fazio}

\affiliation{NEST-INFM $\&$ Scuola Normale
                Superiore, I-56126 Pisa, Italy}

\begin{abstract}
In this paper we review several aspects of mesoscopic hybrid superconducting systems. In particular we consider charge and heat transport properties in hybrid superconducting-metal structures and the effect of charging energy in superconducting nanostructures.
\end{abstract}

\maketitle

\tableofcontents

\section{Introduction}
Mesoscopic physics concerns with the properties of small systems with sizes 
in the range from of a few nanometers to micrometers, at low temperatures typically
below 1 K. The constant progress in nano-fabrication techniques allows for a 
controlled realization of these structures and a consequent increasing interest
in this physics. 
Hybrid superconducting systems are mesoscopic devices constituted by normal and superconducting parts in electrical contact.
Characteristic of superconductivity is the macroscopic phase coherence of 
the order parameter and the supercurrent flow. On the one hand, superconductivity adds 
new degrees of freedom and makes the physics of mesoscopic systems richer.
On the other hand, superconducting properties are deeply influenced by mesoscopic effects as it is the case of charging effects in small superconducting junctions. 

This brief introduction makes evident that the field of hybrid mesoscopic superconducting systems is vast and diversified.
It is our intention to give a general overview of the field, pointing to the relevant references for a more detailed analysis, and to keep the presentation as simple as possible.
The paper is divided in five sections (including this introduction) which are almost independent among themselves.
We begin, in Section II, by introducing the basic properties of hybrid systems whose electrical transport is dominated by the conversion of normal current into supercurrent which takes place at the interface.
We give, for educational purposes, an introduction to the technique of the Bogoliubov-de Gennes equations combined to the scattering approach for the study of transport in hybrid systems. 
The complementary approach using quasiclassical Green functions is then briefly introduced, however we refer to the existing literature for a more detailed presentation.
In Section III we introduce the heat-transport properties of hybrid systems in order to discuss the effectiveness of such systems as electronic coolers.
In Section IV we provide a presentation on the superconducting properties in 
small systems where charging effects are dominant.

Basics in superconductivity and Josephson physics can be found in the books by 
Tinkham~\cite{tinkham96} and by Barone and Patern\`o~\cite{barone82}. Since many ideas 
discussed here belong to the field of more traditional mesoscopic physics we refer
for these topics to the books by Beenakker and van Houten~\cite{beenakker91} and Imry~\cite{imry97} and
to the conference proceedings~\cite{karlsruhe94,curacao97}. Excellent introductions to 
single charge tunneling can be found in Refs.~\cite{averin,grabert,schoen}. We also refer to the recent reviews on hybrid structures~\cite{raimondi} and ultra-small superconductors~\cite{vondelft} for a more complete description of these topics and a more complete list of references.

\section{Hybrid systems}

\subsection{Andreev reflection and proximity effect}

A fascinating aspect of mesoscopics is the possibility of fabricating
hybrid nanostructures formed from combinations of normal
conductors (N) and superconductors (S).
The interest for these systems stems from the fact that the nature of
charge transport is dramatically different in the normal and in the
superconducting constituents of the heterostructure, giving rise to
a rich variety of effects.
While the charge current in N is carried by quasi-particle
excitations (electrons in a metal), at sufficiently low energy (e.g. temperature,
applied voltage, frequency,...) the
current in S is carried by the superconducting condensate, which is a
many body ground state property of the system, and
flows without dissipation.

\begin{figure}
\begin{center}
\epsfig{figure=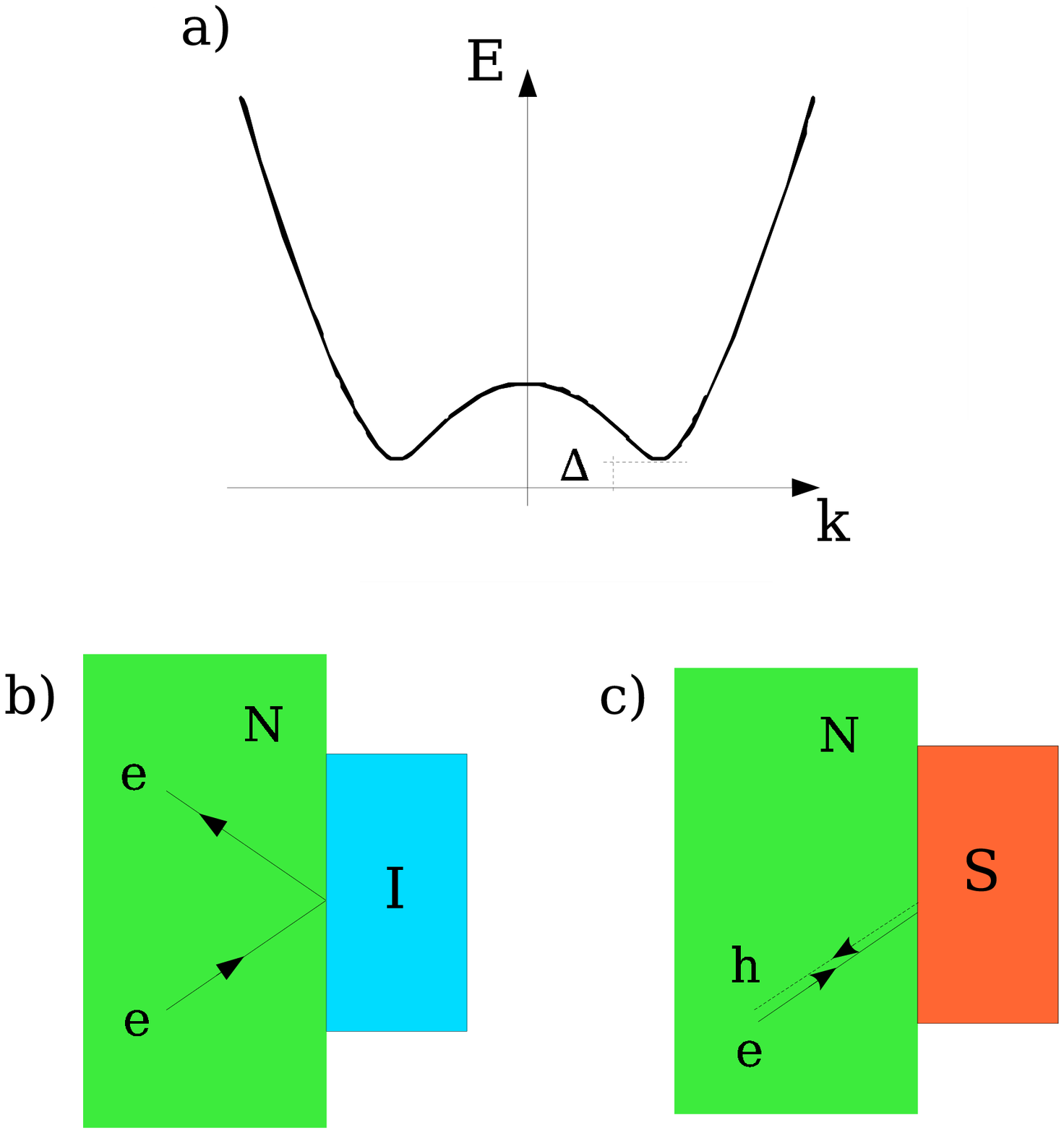,width=0.6 \textwidth}
\end{center}
\caption{a) Energy spectrum of quasi-particle excitations in a superconductor: quasi-particle energies $E$ are plotted as a function of their wave-vectors $k$. Superconducting correlations give rise to a gap of amplitude $|\Delta|$ above the Fermi level set by the horizontal axis. Typical values of $\Delta$ go from hundreds of $\mu eV$ to a few $meV$.
b) In a normal reflection process, which occurs for example at an interface with an
insulator (I), an electron is scattered back into the normal metal conserving the
component of the momentum parallel to the interface.
c) In an Andreev reflection, occurring at a NS interface, the incoming electron is
reflected back as a hole. In this case the component of the momentum parallel to
the interface is reversed, so that, at the Fermi energy, the reflected hole
retrace back the trajectory of the impinging electron.}
\label{Andreevreflection}
\end{figure}

The matching between these two different charge transport mechanisms can
be well understood for phase-coherent structures in terms of the so-called
{\em Andreev reflection} process \cite{andreev64}.
This can be introduced starting from the microscopic BCS theory of
superconductivity \cite{BCS}.
In the presence of boundaries and non-uniformities it is convenient to
perform an approximate Fourier transform of the BCS Hamiltonian.
Making use of the Hartree approximation, one arrives to an effective
single-particle Hamiltonian which leads to the Bogoliubov-de Gennes
equation \cite{BdG}
\begin{equation}
\left( \begin{array}{cc} H_0 & \Delta \\ \Delta^* & -H_0^*
\end{array}\right)
\left( \begin{array}{c}u\\v
\end{array}\right)=
E\left( \begin{array}{c}u\\v
\end{array}\right)
\label{BdGequation}
\end{equation}
for the coherence factors $u$ and $v$, where the energy $E$ is measured from
the condensate chemical potential.
This is a matrix eigenvalue equation which contains the single-particle
Hamiltonian $H_0$ describing electrons in the superconductor in the
absence of the attractive potential, and the time-reversal of such a Hamiltonian
($-H_0^*$), which describes the hole degree of freedom.
The off-diagonal term $\Delta$, which introduces a coupling
between electrons and holes, is given by the product of the point-like
electron-electron attractive potential and the pairing amplitude describing
superconducting correlations among electrons. The pairing amplitude, proportional 
to the anomalous average $\langle \psi \psi \rangle$ can be understood as being 
the wave function of the Cooper pairs.
Since $\Delta$, which is the superconducting order parameter, is determined by the coherence
factors, it should be calculated self-consistently.
In the absence of superconductivity ($\Delta=0$) the system is described in terms of
decoupled electrons and holes, the latter being characterized by having group velocity
and wave-vector in opposite directions, as well as opposite charge and spin with
respect to electrons.
When an electronic attractive interaction is present ($\Delta\ne 0$), electron and hole
degrees of freedom get mixed forming particle-like and hole-like
quasi-particles.
As a result, a gap of amplitude $|\Delta|$ opens up in the energy spectrum
(see Fig. \ref{Andreevreflection}a)
forbidding quasi-particles excitations  with energy inside the gap to propagate
in the superconductor.

Let us now consider a piece of normal metal in contact with a superconductor
and assume that electrons of a given energy $E<|\Delta|$ are injected from
the normal part toward the NS interface.
As explained above, such electrons cannot propagate through the superconductor,
in fact they are allowed to
penetrate as an evanescent wave up to a depth of the order of
the BCS superconducting coherence length $\xi$.
They can however undergo two different processes.
In the first one, {\em normal reflection},
electrons are reflected back into the normal slab and do not contribute to
the charge current (see Fig. \ref{Andreevreflection}b). The second one is made possible by the electron-hole
coupling term of the Bogoliubov-de Gennes Hamiltonian and consists in the
coherent evolution of the incoming electron into a retro-reflected hole
(see Fig. \ref{Andreevreflection}c).
This constitutes the {\em Andreev reflection} process, which is indeed
responsible for charge
transport, and corresponds to the transfer of a pair of electrons from
the normal side of the interface to the superconducting condensate.
Noteworthy, each Andreev reflection process contributes of $2e$ to the charge current.
It is important to notice that Andreev reflection does not contribute to the
transport of energy, since no excitations are transferred into the
superconductor. While providing good charge conductivity, NS interfaces prove to 
be good thermal insulators.

Andreev reflection also accounts for the influence that the superconductor exerts
on the normal conductor, known as {\em proximity effect}.
In much the same way as the quasi-particle excitations penetrate the superconductor,
the superconductor
pairing amplitude also leaks out into the normal side of NS contact over a large distance,
of the order of the superconducting coherence length, even in the absence of an
attractive interaction. As a result, partial superconducting properties are induced
into the normal conductor.

In the next Section (\ref{deriveBdG}) we will give a derivation (following Ref.~\cite{BdG}) of
the Bogoliubov-de Gennes equation, while 
in Section \ref{solveBdG} we shall solve such an equation for different systems.
In the following sections we shall describe the two most widely used theoretical approaches
applied to study hybrid systems, namely the scattering formalism
(Section \ref{Scattering}), which we will analyze in details, and the quasi-classical
Green's function theory (Section \ref{Green}).
In Section \ref{Results} we shall review the most important results attained in
the field.

\subsection{Derivation of the Bogoliubov-de Gennes equation}
\label{deriveBdG}

As mentioned above, the Bogoliubov-de Gennes equation is derived from the BCS theory.
In this section we show how to derive such an equation in the general case where an exchange field
is present. This will be useful for studying hybrid structures in which ferromagnets
are present too.
It is convenient to make use of the
Bogoliubov's self-consistent field method \cite{BdG,bogoliubov59}.
The starting point is the following Hamiltonian in the second quantization form:
\begin{equation}
\hat{H}=  \int d\vec{r} ~\left( \hat{\psi}^\dag_\uparrow (\vec{r}),
\hat{\psi}^\dag_\downarrow (\vec{r}) \right)
{\cal H}(\vec{r}) \left( \begin{array}{c} \hat{\psi}_\uparrow (\vec{r}) \\
\hat{\psi}_\downarrow (\vec{r})
\end{array} \right)
 -\frac{V}{2} \sum_{\sigma\sigma'}
\int d\vec{r} ~\hat{\psi}^\dag_\sigma (\vec{r}) \hat{\psi}^\dag_{\sigma'}
(\vec{r}) \hat{\psi}_{\sigma'} (\vec{r}) \hat{\psi}_{\sigma} (\vec{r}) ~,
\label{ham}
\end{equation}
where the second term accounts for the electron-electron coupling, which is assumed
to be a point-like, two-particle interaction (introduced for the first time by
Gorkov \cite{gorkov58}) and independent of spin,
therefore characterized by a single coefficient $V$.
In Eq. (\ref{ham}),
\begin{equation}
{\cal H}(\vec{r})= \left[ \frac{1}{2m} \left( -i\hbar \vec{\nabla} -\frac{e}{c}
\vec{A}(\vec{r})
\right)^2 +V_0(\vec{r}) -\mu \right] \underline{\underline{1}} -
\frac{\hbar e}{2mc} \left( \underline{\underline{\vec{\sigma}}} \cdot \vec{M}
\right) 
\label{lham}
\end{equation}
is the single-particle Hamiltonian in the presence of an exchange field (second
term). 
$\vec{A}(\vec{r})$ is the vector potential, $V_0(\vec{r})$ is the
normal potential, $\mu$ is the chemical potential, $\vec{M}$ is the internal
mean field due to the exchange interaction and $U_0(\vec{r})$ is the
periodic potential due to the ions in the crystal.
$c$ is the speed of light,
\underline{\underline{1}} is the 2$\times$2 unity matrix and
$\underline{\underline{\vec{\sigma}}}$ is the vector of Pauli matrices.
Before proceeding, we want to remark that ferromagnetism is
introduced according to the Stoner model (see Refs. \cite{Zim,Ston1,Ston2,Enz}).
In this model a molecular field, due to the permanent magnetization present in
a ferromagnet, replaces the external magnetic field in the Zeeman energy, and it
is added to the single-particle energy.
Such a molecular field can be obtained calculating the exchange interaction
in a system of spins using the mean field approximation (see Ref. \cite{Amik}).
Such a molecular field can be treated as an adjustable
parameter to fit band structure calculations.

The operators $\hat{\psi}_\sigma (\vec{r})$ and
$\hat{\psi}^\dag_\sigma (\vec{r})$ are, respectively, destruction and creation
field operators for a
particle of spin $\sigma$ at the point $\vec{r}$, which obey the Fermi
commutation relations
\begin{eqnarray}&
\hat{\psi}_\sigma (\vec{r}) \hat{\psi}_{\sigma'} (\vec{r'}) +
\hat{\psi}_{\sigma'} (\vec{r'}) \hat{\psi}_\sigma (\vec{r}) =0 \\&
\hat{\psi}^\dag_\sigma (\vec{r}) \hat{\psi}_{\sigma'} (\vec{r'}) +
\hat{\psi}_{\sigma'} (\vec{r'}) \hat{\psi}^\dag_\sigma (\vec{r})
=\delta_{\sigma\sigma'} \delta (\vec{r}-\vec{r'}) ~.
\end{eqnarray}
We now seek an approximate solution of the many body problem in terms of an
effective single-particle Hamiltonian $\hat{H}_{eff}$.
According to the Hartree approximation (see Ref. \cite{Gross}), such a
Hamiltonian describes a particle which moves in a mean-field potential produced
by all other particles.
Using the variational principle, it can be shown that
\begin{eqnarray}&&
\hat{H}_{eff} =\int d\vec{r} \left( \hat{\psi}^\dag_\uparrow (\vec{r})
\hat{\psi}^\dag_\downarrow (\vec{r}) \right) \left[ {\cal H}(\vec{r})
+\left( \begin{array}{cc} {\cal U}^{\uparrow}(\vec{r}) &0\\0& {\cal
U}^{\downarrow} (\vec{r}) \end{array} \right) \right]
\left( \begin{array}{c}
\hat{\psi}_\uparrow (\vec{r}) \\
\hat{\psi}_\downarrow (\vec{r})
\end{array} \right) + \nonumber \\ &&
+\int d\vec{r} \left[ \hat{\psi}^\dag_\uparrow (\vec{r})
\hat{\psi}^\dag_\downarrow (\vec{r}) ~\Delta(\vec{r}) +\hat{\psi}_\downarrow
(\vec{r}) \hat{\psi}_\uparrow(\vec{r}) ~\Delta^\ast(\vec{r}) \right] ~,
\label{Heffe}
\end{eqnarray}
where
${\cal U}^{\uparrow}(\vec{r})$, ${\cal U}^{\downarrow}(\vec{r})$ and
$\Delta(\vec{r})$ are effective potentials defined by
\begin{eqnarray}&&\label{bcs-u-up}
{\cal U}^\uparrow (\vec{r})= -V \langle\hat{\psi}^\dag_\downarrow(\vec{r})
\hat{\psi}_\downarrow(\vec{r}) \rangle  ~, \\&&
\label{bcs-u-down}
{\cal U}^\downarrow (\vec{r})= -V \langle\hat{\psi}^\dag_\uparrow(\vec{r})
\hat{\psi}_\uparrow(\vec{r}) \rangle
\end{eqnarray}
and  
\begin{equation}
\label{bcs-delta}
\Delta (\vec{r})= -V \langle \hat{\psi}_\downarrow(\vec{r})
\hat{\psi}_\uparrow(\vec{r}) \rangle=
V \langle \hat{\psi}_\uparrow(\vec{r})\hat{\psi}_\downarrow(\vec{r}) \rangle ~.
\end{equation}
${\cal U}^{\uparrow} (\vec{r})$ and ${\cal U}^{\downarrow}(\vec{r})$
are single particle potentials (standard Hartree result for
point-like interactions) and $\Delta (\vec{r})$ is referred to as the
pairing potential.

The key feature of the mean-field theory of superconductivity is that we admit
non-vanishing expectation values for the spin-paired operators
$\hat{\psi}_\uparrow(\vec{r}) \hat{\psi}_\downarrow(\vec{r})$ and
$\hat{\psi}_\downarrow(\vec{r}) \hat{\psi}_\uparrow(\vec{r})$.
If we consider $\hat{H}_{eff}$ in (\ref{Heffe}), we see that,
on the one hand, the terms in ${\cal U}^{\uparrow}$ and ${\cal U}^{\downarrow}$
destroy and create one
electron and therefore conserves the number of particles. On the other hand, the
terms in $\Delta$ increases or decreases the number of particles by 2. This is not
a problem, since $\hat{H}_{eff}$ operates on the BCS wave function
which is not an eigenfunction of the number operator.
Applying the following unitary transformation
\begin{equation}
\begin{array}{c}
\hat{\psi}_\uparrow =\sum_n \left( u_n^{\uparrow}(\vec{r})\hat{\gamma}_{n\uparrow} -
v_n^{\uparrow\ast}(\vec{r}) \hat{\gamma}^\dag_{n\downarrow} \right)\\
\hat{\psi}_\downarrow=\sum_n \left( u_n^{\downarrow}(\vec{r})\hat{\gamma}_{n\downarrow} +
v_n^{\downarrow\ast}(\vec{r}) \hat{\gamma}^\dag_{n\uparrow} \right)
\end{array} ~,
\label{Bogtr}
\end{equation}
$\hat{H}_{eff}$ can be diagonalized in such a way that
\begin{equation}
\hat{H}_{eff}= E_g +\sum_{n\sigma} E_n \hat{\gamma}^\dag_{n\sigma}
\hat{\gamma}_{n\sigma} ~,
\label{haml}
\end{equation}
where $E_g$ is the superconducting ground state energy, $E_n$ is the energy of
the $n$-th quasi-particle excitation and $\hat{\gamma}_{n\sigma}$
($\hat{\gamma}^\dag_{n\sigma}$) is the quasi-particle destruction (creation) operator
satisfying Fermi commutation relations:
\begin{eqnarray}
\left\{ \hat{\gamma}^\dag_{n\sigma}, \hat{\gamma}_{m\sigma'} \right\} &=\delta_{nm}~
\delta_{\sigma\sigma'} \label{relF1}\\
\left\{ \hat{\gamma}_{n\sigma}, \hat{\gamma}_{m\sigma'} \right\} &=0 ~.
\label{relF}
\end{eqnarray}

The Bogoliubov-de Gennes equation is an eigenvalue equation which allows one to calculate the energy spectrum of the quasi-particles and the coherence factors $u_n^\sigma
(\vec{r})$ and $v_n^\sigma (\vec{r})$:
\begin{equation}
\left( \begin{array}{cccc}
(\overline{H}^\uparrow -h_z) & 0 &0& \Delta(\vec{r})\\
0&(\overline{H}^\downarrow +h_z)&\Delta(\vec{r})&0
\\
0&\Delta^\ast(\vec{r})&(-\overline{H}^\uparrow
+h_z)&0\\
\Delta^\ast(\vec{r})&0&0&(-\overline{H}^\downarrow
-h_z)
\end{array} \right) \cdot
\left( \begin{array}{c} u_n^\uparrow(\vec{r})\\ u_n^\downarrow(\vec{r})\\
v_n^\uparrow(\vec{r})\\ v_n^\downarrow(\vec{r}) \end{array}\right)=E_n
\left( \begin{array}{c} u_n^\uparrow(\vec{r})\\ u_n^\downarrow(\vec{r})\\
v_n^\uparrow(\vec{r})\\ v_n^\downarrow(\vec{r}) \end{array}\right) ~,
\label{questa}
\end{equation}
where $\overline{H}^\sigma (\vec{r})=H_0+{\cal U}^\sigma (\vec{r})$, with
\begin{equation}
H_0=\left[ -\frac{\hbar^2}{2m} \vec{\nabla}^2
+V_0(\vec{r}) -\mu \right] ~,
\end{equation}
\begin{equation}
h_z=\frac{\hbar e}{2mc}M_z~,
\label{hexch}
\end{equation}
having set $\vec{A}=0$ and $M_x=M_y=0$.

We shall refer to the matrix Hamiltonian in (\ref{questa}) as the Bogoliubov-de
Gennes Hamiltonian $H_{BG}$.
If we identify the top-left 2$\times$2 block of such a Hamiltonian
with the particle Hamiltonian $H_p$ and the bottom-right
2$\times$2 block with the hole Hamiltonian $H_h$, we see
that the equality $H_h= -H_p^\ast$ holds.

So far, we have diagonalized $\hat{H}_{eff}$ using the transformation
(\ref{Bogtr}), which was proposed independently by Bogoliubov \cite{Bogo} and
Valatin \cite{Vala}, known as Bogoliubov-Valatin transformation.
$\hat{H}_{eff}$
is now written in terms of quasi-particle operators $\hat{\gamma}$,
$\hat{\gamma}^\dag$, which greatly simplifies calculations.
Although the Bogoliubov-de Gennes equation (\ref{questa}) does not determine the
potentials
${\cal U}^\uparrow (\vec{r})$, ${\cal U}^\downarrow (\vec{r})$
and $\Delta(\vec{r})$, they can be fixed
self-consistently.
By replacing the field operators in the definition of the above potentials (\ref{bcs-u-up}),
(\ref{bcs-u-down}), (\ref{bcs-delta}) and using the Bogoliubov-Valatin
transformation (\ref{Bogtr}) we obtain
\begin{equation}
{\cal U}^\uparrow (\vec{r})=-V\sum_n \left[ |u_n^\downarrow (\vec{r})|^2
 f_n +
|v_n^\downarrow (\vec{r})|^2 (1-f_n) \right]~,
\label{u1}
\end{equation}
\begin{equation}
{\cal U}^\downarrow (\vec{r})=-V\sum_n \left[ |u_n^\uparrow (\vec{r})|^2
 f_n +
 |v_n^\uparrow (\vec{r})|^2 (1-f_n)
\right]
\label{u3}
\end{equation}
and
\begin{equation}
\Delta(\vec{r})=V\sum_n \left[ u_n^{\downarrow} v_n^{\uparrow\ast} ~(1-f_n)-
v_n^{\downarrow\ast}u_n^{\uparrow}~f_n \right] ~,
\label{u2}
\end{equation}
where
\begin{equation}
f_n=\frac{1}{e^{\frac{E_n}{K_{\text{B}}T}} +1} ~.
\end{equation}
Equations (\ref{u1}), (\ref{u3}) and (\ref{u2}) constitute the self-consistent
equations for ${\cal U}^{\uparrow}(\vec{r})$, ${\cal U}^{\downarrow}(\vec{r})$
and $\Delta(\vec{r})$.
These ensure that if we solve the Bogoliubov-de Gennes equation (\ref{questa}), then
the value of ${\cal U}^{\uparrow}$, ${\cal U}^{\downarrow}$ and $\Delta$ which
is calculated from the solutions of (\ref{u1}), (\ref{u3}) and (\ref{u2}) is
equal to the initial value of ${\cal U}^{\uparrow}$, ${\cal U}^{\downarrow}$
and $\Delta$.

There is an important distinction between ${\cal U}^{\uparrow}(\vec{r})$ and
${\cal U}^{\downarrow}(\vec{r})$, on one side, and $\Delta(\vec{r})$ on the
other.
The Hartree potentials ${\cal U}^{\uparrow}(\vec{r})$ and ${\cal
U}^{\downarrow}(\vec{r})$ come from a sum
involving all states below the Fermi level and hence are nearly temperature
independent and can be approximated by the Hartree potentials calculated in the
normal state. However, the pair potential $\Delta(\vec{r})$ is a sum of terms of
the form $u_n^\sigma (\vec{r})~v_n^{\sigma'\ast}(\vec{r})$ which are non-zero
only in the
neighborhood of the Fermi surface. For this reason $\Delta(\vec{r})$ is a
strong function of temperature.

\subsection{Solutions of the Bogoliubov-de Gennes equation}
\label{solveBdG}
In this section we discuss the solutions of the Bogoliubov-de Gennes equation
(\ref{questa}) in heterostructures
containing normal metals, ferromagnets and superconductors. 
In the following
we provide, first, the solution in the case of a homogeneous and clean
superconductor (where $\vec{h}=0$) and finally a
solution for a ferromagnet/superconductor interface.

\subsubsection{Superconductor}
\label{sup-BdG}

\begin{figure}
\begin{center}
\epsfig{figure=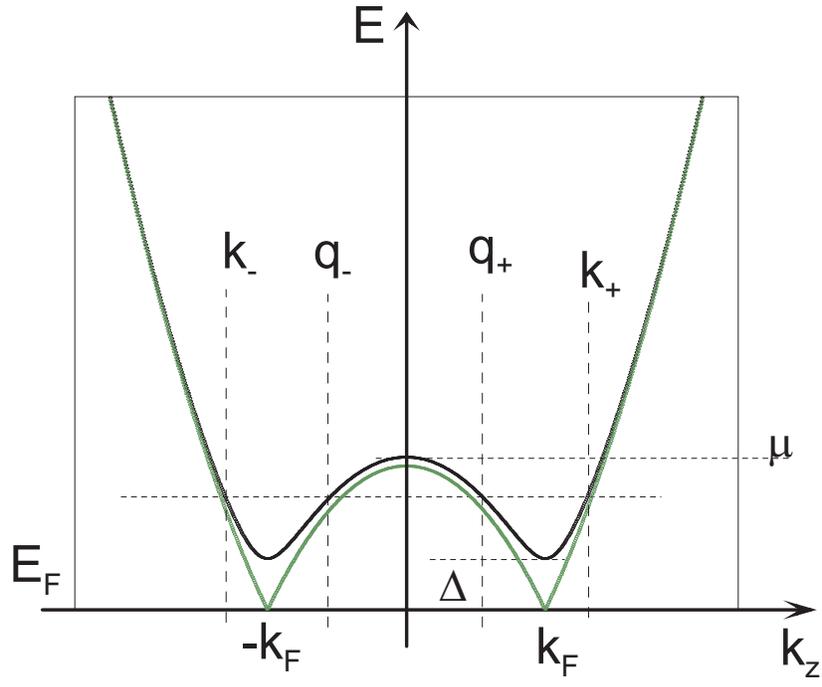,width=0.6 \textwidth}
\end{center}
\caption{Dispersion curve relative to a uniform superconductor for
$\Delta=1/3\mu$ (solid line) and $\Delta=0$ (grey line), taking
$\Delta=1$, $\mu=3$ and $k_x=k_y=2\pi/20$. For $\Delta\neq 0$ an energy gap of
amplitude $|\Delta|$ appears in the spectrum.}
\label{noSO.s.cont}
\end{figure}

We now consider another limiting case: all potentials are set to zero apart
from $\Delta\neq 0$. The Bogoliubov-De
Gennes equation  (\ref{questa}) again decouples into two sets of equations, one
relative to spin $\uparrow$ particles and spin $\downarrow$ holes
\begin{equation}
\left( \begin{array}{cc}
H & \Delta \\ \Delta^\ast & -H
\end{array} \right)\left( \begin{array}{c} u(\vec{r}) \\ v(\vec{r})
\end{array} \right) = E\left( \begin{array}{c} u(\vec{r}) \\ v(\vec{r})
\end{array} \right)
\label{bcs-sup-BG}
\end{equation}
and another, equivalent, relative to spin $\downarrow$ particles and spin
$\uparrow$ holes. Considering plane wave solution of the form
\begin{equation}\left( \begin{array}{c} u(\vec{r}) \\ v(\vec{r})
\end{array} \right) =\left( \begin{array}{c} \psi \\ \phi \end{array} \right)
e^{i\vec{k}\cdot\vec{r}}
\end{equation}
one finds the following dispersion relation:
\begin{equation}
E=\pm \sqrt{(k^2-\mu)^2 +|\Delta|^2} ~,
\end{equation}
which is plotted in Fig. \ref{noSO.s.cont} in the direction $k_z$ when 
$|\Delta|=\mu/3$
(solid line) and $|\Delta|=0$ (gray line). We can notice that there is an
energy gap in the energy spectrum for non-zero $\Delta$ as expected.
It is useful to write down the solution for $k_z$ at a given energy $E$:
\begin{equation}
k_z=\pm \sqrt{\overline{\mu}\pm \sqrt{E^2-|\Delta|^2}} ~,
\end{equation}
where $\overline{\mu}=\mu-k_x^2-k_y^2$.
This tells us that for an energy $E$ above the gap there are four possible
solutions denoted by $k_+$, $k_-$, $q_+$ and $q_-$ in Fig. \ref{noSO.s.cont}. We shall
denote excitations with wave-vectors $k_\alpha$ as particle-like excitations and
those with wave-vectors $q_\alpha$ to be hole-like excitations. As we can see
from Fig. \ref{noSO.s.cont}, this choice means that particle-like excitations have their
group velocity parallel to their momenta, whereas hole-like excitations have
their group velocity anti-parallel to their momenta. This choice is made
because the wave function associated with wave-vector $k_\alpha$ ($q_\alpha$) is
predominantly particle-like (hole-like). This can be seen by substituting
$k_{\alpha}$ and $q_{\alpha}$ in the amplitudes $\psi$ and $\phi$, solutions of
equation (\ref{bcs-sup-BG}):
\begin{equation}
\psi=\frac{e^{i\varphi/2}}{\sqrt{2}} \sqrt{1+\frac{\sqrt{E^2-|\Delta|^2}}{E}}~,
\end{equation}
\begin{equation}
\phi=\frac{e^{-i\varphi/2}}{\sqrt{2}} \sqrt{1-\frac{\sqrt{E^2-|\Delta|^2}}{E}}~,
\end{equation}
where $\varphi$ is the phase of the order parameter.
Note also that since only
the absolute values of the coherence factors $\psi$ and $\phi$ are fixed
($|\psi|^2+|\phi|^2=1$), there is an arbitrary choice in where to put the phase
of the order parameter.

\subsubsection{Ferromagnet/Superconductor (F/S) interface}
\label{BdGFS}

In this section we solve analytically the Bogoliubov-De Gennes equation for a
F/S interface. We shall prove that the Andreev reflection amplitude $r_a$ is
suppressed when we are in the presence of an exchange field
$\vec{h}$.
Before proceeding, note that, for a ballistic N/S interface, $r_a$ was
first calculated by Blonder, Tinkham and Klapwijk (BTK) in Ref. \cite{BTK82}.
$r_a$ was determined as a function of the energy of the quasi-particles and as a
function of the strength of a barrier potential at the interface.
The generalization of the BTK calculation to the case of a F/S interface was
first derived in Ref. \cite{dejong95}.

For simplicity, let us consider a one-dimensional structure in which all
potentials are set to zero apart from
$\vec{h}=h~\theta(-x)~\hat{z}$ and $\Delta(x)=\Delta_0~\theta(x)$, where $h$ and
$\Delta_0$ are constants. In this case, the Bogoliubov-De Gennes equation
(\ref{BdGequation}) can be decoupled into two equivalent equations, the first of which
reads
\begin{equation}
\left( \begin{array}{cc} (-\partial_x^2 -\mu-h) & \Delta(x) \\
\Delta^\ast(x) & (\partial_x^2 +\mu-h) \end{array} \right) \left(
\begin{array}{c} u^\uparrow(x)\\ v^\downarrow(x) \end{array} \right)=
E\left( \begin{array}{c} u^\uparrow(x)\\ v^\downarrow(x) \end{array} \right) ~.
\label{sta}
\end{equation}
We now consider plane wave solutions of equation (\ref{sta}) and we solve the
scattering problem. In the left-hand ($x<0$) ferromagnetic region the wave
function $\psi_L(x)$ produced by a source of right-going spin $\uparrow$
particles of unit flux at energy $E$ can be written as
\begin{equation}
\psi_L(x)=\left( \begin{array}{c} u^\uparrow(x)\\ v^\downarrow(x) \end{array}
\right)= \left( \begin{array}{c} \frac{e^{ikx}}{\sqrt{\nu_k}} +r_0~
\frac{e^{-ikx}}{\sqrt{\nu_k}} \\
r_a~\frac{e^{iqx}}{\sqrt{\nu_q}}
\end{array} \right) ~,
\end{equation}
where $k=\sqrt{E+\mu+h}$,
$q=\sqrt{-E+\mu-h}$. $\nu_k=2k$ and $\nu_q=2q$ are the group velocities
relative, respectively, to the particle wave-vector $k$ and to the hole
wave-vector $q$.
$r_0$ is the normal reflection amplitude, whereas $r_a$ is the Andreev
reflection amplitude, which corresponds to the reflection of an incoming spin
$\uparrow$ particle into a spin $\downarrow$ hole. The Andreev reflection
process \cite{andreev64} consists of the coherent evolution of a particle-like
excitation into a hole-like excitation.
In the right-hand ($x>0$) superconducting region the wave
function $\psi_R(x)$ is
\begin{equation}
\psi_R(x)=\left( \begin{array}{c} u^\uparrow(x)\\ v^\downarrow(x) \end{array}
\right)= t_0 \left( \begin{array}{c} \psi^+\\ \phi^+ \end{array}\right)
\frac{e^{ik_+x}}{\sqrt{\nu_{k_+}}} +t_a \left( \begin{array}{c} \psi^-\\ \phi^-
\end{array}\right) \frac{e^{-iq_+x}}{\sqrt{\nu_{q_+}}} ~,
\label{pii}
\end{equation}
where $\psi^\pm$ and $\phi^\pm$ are the coherence factors, solutions
of the Bogoliubov-de Gennes equation (\ref{bcs-sup-BG}) for a superconductor
(see sub-Section \ref{sup-BdG}), given by
\begin{equation}
\left( \frac{\psi}{\phi} \right)^\pm=\frac{\Delta^\ast}{E\pm
\sqrt{E^2-\Delta_0^2}}
\label{3.101}
\end{equation}
and $|\psi^\pm|^2+|\phi^\pm|^2=1$.
$t_0$ and $t_a$ are, respectively, normal and
Andreev transmission amplitudes.
In (\ref{pii}) we have defined:
\begin{equation}
k_+=\sqrt{\mu+\sqrt{E^2-\Delta_0^2}} ~,
\end{equation}
\begin{equation}
q_+=\sqrt{\mu-\sqrt{E^2-\Delta_0^2}} ~,
\end{equation}
\begin{equation}
\nu_{k_+}=\frac{\partial E}{\partial k_z}
=\frac{2k_+ (k_+^2-\mu)}{\sqrt{(k_+^2-\mu)^2+\Delta_0^2}}
\end{equation}
and similarly for $\nu_{q_+}$.
Since we are interested in the subgap solutions ({\it i. e.} when $E<\Delta_0$), both $k_+$
and $q_+$ are complex and $\psi_R(x)$ is an evanescent wave.
The length $\xi$ over which the decay of $\psi_R(x)$ occurs can be defined by
$\mbox{Re} [ik_+ \xi]=-1$, so that $\xi=1/\mbox{Im} [k_+]$. $\xi$ is known as
the superconducting coherence length and for $E=0$ one has $\xi\sim
\frac{1}{\Delta}$.
It is therefore
useful to define: $k_+=\sqrt{\mu+i\eta}$ and $q_+=\sqrt{\mu-i\eta}$ with
$\eta=\sqrt{\Delta_0^2-E^2}$.
The scattering problem is solved once $r_0$, $r_a$, $t_0$ and $t_a$ are
calculated. This can be done by matching the wave functions $\psi_L$ and
$\psi_R$ and their derivatives at the interface ($x=0$).
We find
\begin{equation}
r_a= -\left( \frac{\nu_q}{\nu_k} \right) \frac{2k (q_++k_+)}
{(\epsilon+k_+)(\phi-q_+) \left( \frac{\psi}{\phi} \right)^+ -
(\epsilon-q_+)(\phi+k_+) \left( \frac{\psi}{\phi} \right)^-} ~.
\label{sysra}
\end{equation}
In the case where $H=0$ and $E=0$, in the limit of small $\Delta_0/\mu$,
(\ref{sysra}) reduces to
\begin{equation}
r_a=2ie^{i\varphi} \frac{\left( \mu \sqrt{\mu^2-h^2}\right)^{1/2}}{\mu+\
\sqrt{\mu^2-h^2}} ~,
\end{equation}
in such a way that the Andreev reflection
probability $R_a=|r_a|^2$ is a decreasing function of $h$.
For small values of $h$, $r_a$ at $E=0$ can be approximated by
\begin{equation}
r_a \simeq i~e^{i\varphi}
\label{ra_lim}
\end{equation}
like in the N/S case.
In real hard ferromagnets, like Co, the value of the exchange field, although
high with respect to other ferromagnets, is one order of magnitude smaller
than the Fermi energy, and therefore (\ref{ra_lim}) holds.
In 3-dimensions, however, the picture changes since $\mu$ has to be replaced by
$\mu -E_n$, where $E_n$ is the transverse kinetic energy relative to the $n$-th
longitudinal mode.
For quasi-particles approaching the interface at large angles, $\mu -E_n$ can
become comparable to or smaller than $h$ and, as a result, the suppression of
$R_a$ is enhanced with respect to the 1-dimensional case.

To conclude we remark, firstly, that $r_a$ contains the phase of the
superconducting order parameter $\varphi$. It is this which gives rise to
interference phenomena for systems containing more than one superconducting
interface with different phases. Secondly, Andreev reflection is the
process by which electrical current can flow across a
normal-metal/superconductor interface in the subgap regime.
Since the change in momentum required for the Andreev process ($\delta
p=\hbar(k-q)\simeq \hbar k_F E/\mu$) is much smaller than the change in momentum
required for a normal reflection process ($\delta p=2\hbar k \simeq 2\hbar
k_F$), Andreev reflection is strongly favored at a clean
normal-metal/superconductor interface for subgap energies.
At the Fermi energy ($E=0$) Andreev reflected hole retrace the trajectory of the incoming electrons ({\it retroreflection}).
During this process a charge of $2e$ is deposited into the superconducting
condensate in the form of a Cooper pair, which would be carried away by a
supercurrent.
For energies greater that $|\Delta|$ normal transmission is dominant for clean
normal-metal/superconductor interfaces.

\subsection{Scattering theory}
\label{Scattering}
Electronic transport through a phase-coherent conductor can be studied in the same way
as wave propagation into a scatterer.
A conductor can be modeled as a scattering region connected to external leads which
act as waveguides carrying a current of quasi-particles originating in external reservoirs.
Scattering is assumed to be elastic and
all inelastic processes occur in such reservoirs where quasi-particles are distributed
in energy according to the Fermi-Dirac function
$f(E)=[\exp[(E-\mu)/k_BT]+1]^{-1}$ and
characterized by a chemical
potential $\mu=eV$ and a temperature $T$, $e$ being the electronic charge.
The conductance $G$ of
completely normal structures can be expressed in terms of scattering
probabilities \cite{landauer70,buttiker85,buttiker86,buttiker88}
yielding the Landauer-B\"uttiker formula:
\begin{equation}
G=\frac{2e^2}{h} \text{Tr} [t^\dagger t]
\label{landauer}
\end{equation}
for a 2-terminal system at zero temperature in the linear response regime.
In Eq. (\ref{landauer}) $e^2/h$ is the quantum of conductance, the factor 2
accounts for the twofold spin degeneracy and $t$ is the transmission amplitude matrix
for an electron to propagate through the one-dimensional conductor.

For hybrid systems the scattering theory was first applied \cite{SNS}
to a NIS interface
(I being an insulating barrier of arbitrary strength) in Ref. \cite{BTK82}.
The total current $I_{\text{NS}}$ was calculated in the normal side of the junction in terms
of the difference between incoming and outgoing quasi-particle distribution functions yielding
the so-called BTK formula:
\begin{equation}
I_{\text{NS}}=2N(0)ev_{\text{F}} \mathcal{A}\int_{-\infty}^{\infty}
[f(E-eV)-f(E)][1+A(E)-B(E)]dE ~,
\label{BTKcurrent}
\end{equation}
where $A(E)$ and $B(E)$ are the energy-dependent Andreev and normal reflection probabilities,
respectively, and $V$ is the bias voltage applied to the junction.
In Eq. (\ref{BTKcurrent}) $N(0)$ is the density of states at the Fermi energy,
$v_\text{F}$ is the Fermi velocity, $\mathcal{A}$ is the effective cross-sectional area
and $f$ is the Fermi distribution function.
$A$ and $B$ were determined by imposing matching conditions to the wavefunctions, solution
of the Bogoliubov-de Gennes
equation, at the N and S side assuming a $\delta$-like potential located at the interface.
For low transmitting interfaces $I_{\text{NS}}$ reproduces the conventional transfer Hamiltonian
result for which the current is proportional to the density of states of the
superconductor and to the transparency of the interface (Andreev processes are completely
neglected).
Namely $I_{\text{NS}}$ is vanishingly small for sub-gap voltages and
presents a pronounced peak at $eV=\Delta$ which rapidly decays asymptotically reaching the
normal state current value.
For highly-transmitting interfaces the I/V curve presents a large sub-gap current, due to
Andreev reflection,
which takes its maximum value when $eV$ approaches $\Delta$ and thereafter decreasing to
the normal state value.
In addition it is also possible to determine that the length at which the
quasi-particle evanescent wave penetrates into the superconductor is equal to
$\hbar v_{\text{F}}/2\Delta$ at energies close to the Fermi energy,
i.e. of the same order of the BCS superconducting coherence length
$\xi=\hbar v_{\text{F}}/\pi\Delta$.

A more general treatment generalizing the Landauer-B\"uttiker formulae to superconducting
systems was developed in Refs.\cite{lambert91,takane92,lambert94}.
It accounts for generic multi-terminal, multi-channel hybrid structures,
even in the presence of ferromagnetism and spin-flip scattering.
As mentioned before, in the absence of inelastic scattering
(although a generalization to include this is available \cite{buttiker86a}), dc
transport is determined by the quantum mechanical scattering matrix $S(E,{\cal H})$,
which yields scattering properties at energy $E$, of a
phase coherent structure described by a Hamiltonian ${\cal H}$.

Consider a scattering region connected to external leads which carry a current
of quasi-particles originating in external reservoirs at chemical potential
$\mu_i=ev_i$.
In the presence of Andreev scattering, current-voltage relations for a
phase-coherent scatterer connected to normal reservoirs
were first derived in
Ref. \cite{lambert91}.
If the structure is
connected to external reservoirs by current-carrying leads with
open scattering channels
labeled by a set of quantum
numbers $n$, then the S-matrix elements are defined through the relation:
\begin{equation}
\hat{b}_n(E)=\sum_{n'}S_{n,n'}(E,{\cal H}) ~\hat{a}_{n'}(E)\\
\end{equation}
which connects the second quantization operators $\hat{b}_n$ of quasi-particles
leaving the scattering region
through channel $n$ to the operators $\hat{a}_{n'}$ of quasi-particles entering
the scattering region through channel $n'$.
Both $\hat{a}_n$ and $\hat{b}_n$ satisfy anticommutation relations.
The S-matrix satisfies the unitarity condition
$S^{-1}(E,{\cal H})= S^\dagger(E,{\cal H})$,
due to quasi-particle conservation,
and the time-reversibility condition $S^t(E,{\cal H})= S(E,{\cal H}^\ast)$.
In the presence of superconductivity
it is useful to label the quasi-particle open channels in the leads using
the following set of discrete quantum numbers: $\{i,a,\alpha\,\sigma\}$, where
$\alpha=+1$ characterizes particle-like excitations, $\alpha=-1$ hole-like
excitations, $a$ is the open channel index in lead $i$ and $\sigma=\pm 1$ for spin up (down).
In doing so, the quantity
$P_{i,j}^{\alpha\sigma,\beta\sigma'}(E,{\cal H}) =\sum_{a,b} \left| S_{(i,a),(j,b)}^{\alpha\sigma,\beta\sigma'}
(E,{\cal H}) \right|^2$
is the probability of reflection (if $i=j$) or transmission (if $i\neq j$)
of a spin $\sigma'$ quasi-particle of type $\beta$ in lead $j$ to a spin $\sigma$ quasi-particle of type
$\alpha$ in lead $i$. For $\alpha \neq \beta$ $P_{i,j}^{\alpha\sigma,
\beta\sigma'}$ is
referred to as an Andreev scattering probability, while for $\alpha = \beta$, it
is a normal scattering probability.
In the presence of superconducting leads, we insist that all
superconductors share a common condensate chemical potential $\mu$. This is to
avoid time-dependent order parameter phases varying at the Josephson frequency,
which would render a time-independent scattering approach invalid.
When the energy $E$ is measured with respect to $\mu_\text{S}$, the particle-hole
symmetry 
$S^{\alpha\sigma,\beta\sigma'}_{(i,a),(j,b)} (E,{\cal H})=\alpha\beta \left[
S^{-\alpha\sigma,-\beta\sigma'}_{(i,a),(j,b)} (-E,{\cal H}) \right]^\ast$
is satisfied too.

Because unitarity yields
\begin{equation}
\sum_{\beta,\sigma',j,b} \left| S_{(i,a),(j,b)}^{\alpha\sigma,\beta\sigma'}
(E,{\cal H})
\right|^2 =\sum_{\alpha,\sigma,i,a} \left|
S_{(i,a),(j,b)}^{\alpha\sigma,\beta\sigma'} (E,{\cal H})\right|^2 =1 ~,
\end{equation}
where $i$ and $j$ sum only over leads supporting open channels at energy $E$, we
have
\begin{equation}
\sum_{\beta,\sigma',j}P_{i,j}^{\alpha\sigma,\beta\sigma'} (E,{\cal H}) =
N_i^{\alpha\sigma} (E)
\label{unit1}
\end{equation}
and
\begin{equation}
\sum_{\alpha,\sigma,i}P_{i,j}^{\alpha\sigma,\beta\sigma'} (E,{\cal H}) =
N_j^{\beta\sigma'} (E)
\end{equation}
where $N_i^{\alpha\sigma} (E)$ is the number of open channels for $\alpha$-type
$\sigma$-spin
quasi-particles of energy $E$ in lead $i$.

The total current due to all quasi-particles is the integral
\begin{equation}
I_i=\frac{e}{h} \int_0^\infty dE \sum_{\alpha\sigma} (\alpha) \left[
N_i^{\alpha\sigma}(E) f_i^{\alpha}(E) -\sum_{j\beta\sigma'}
P_{i,j}^{\alpha\sigma,\beta\sigma'}(E) f_j^{\beta}(E) \right] ~,
\label{lambertformula}
\end{equation}
where $f_j^\beta(E)$ is the Fermi distribution of incoming quasi-particles of
type $\beta$ from reservoir $j$ at temperature $T_j$ and chemical potential
$\mu_j$:
\begin{equation}
f_j^\beta (E)=\frac{1}{e^{\frac{E-\beta (\mu_j -\mu)}{k_B T_j}}+1}
\label{Fer}
\end{equation}
and $\mu$ is the chemical potential of the condensate, when superconductors are
present.
At this point it is important to notice that all scattering coefficients can be
computed in the presence of self-consistently determined order parameter
$\Delta$ and normal potentials ${\cal U}^\sigma$. In such a case one must take into
account the effects on the system of changes in the applied potentials $v_j$'s.
Such
changes affect both the normal potential and the superconducting order
parameter. For normal structures, self-consistency in the normal potential has
been considered by B\"uttiker {\it et al.} (Ref. \cite{christen96}) and
it turned out that such effect
can become important at large voltages. Calculations taking into
account a self-consistent order parameter have been carried out for one
dimensional structures in Ref. \cite{martin95,sanchez95,martin96}. These demonstrate that
provided
the currents are low enough with respect to the critical current, the matrix
$P_{i,j}^{\alpha\sigma,\beta\sigma'}$ can remain unchanged, even by the application of finite
voltages of
order $|\Delta|$.

In the linear response limit, i.e. for $(v_j-v)\rightarrow 0$, we
write down $I_i$ in the form:
\begin{equation}
I_i=\sum_{j\sigma} a_{ij}^\sigma (v_j-v) ~.
\label{a}
\end{equation}
with
\begin{equation}
a_{ij}^\sigma= \frac{e^2}{h} \int_{-\infty}^\infty dE \left( -\frac{\partial
f}{\partial E}\right) \sum_{\sigma'} \left[ N_i^{+\sigma'}(E) \delta_{ij}
\delta_{\sigma\sigma'} - P_{ij}^{+\sigma',+\sigma} (E) +
P_{ij}^{+\sigma',-\sigma} (E) \right] ~.
\label{a4}
\end{equation}
At zero temperature, where $-\frac{\partial f}{\partial E}=\delta(E)$,
(\ref{a4}) reduces to
\begin{equation}
a_{ij}^\sigma= \frac{e^2}{h} \sum_{\sigma'} \left[ N_i^{+\sigma'}(0)
\delta_{ij} \delta_{\sigma\sigma'} - P_{ij}^{+\sigma',+\sigma} (0) +
P_{ij}^{+\sigma',-\sigma} (0) \right] ~.
\end{equation}

We conclude by noting that, at finite voltages, the chemical potential of the
condensate $\mu$ should be determined self-consistently by imposing the
conservation of the current $\sum_i I_i=0$.
From the expression of $I_i$ in (\ref{lambertformula}) it is clear that such a
self-consistency condition involves integrals over all incident quasi-particle
energies, which in turns require a knowledge of the S-matrix over a range of
energies.
In the case where
superconductivity is
present in one lead (say $j$) the solution is trivial, since $\mu$ coincides
with one of the voltages $v_j$.
In the general case, however, the solution of the self-consistent equations is
difficult, but for some simple structures may be determined by symmetry
arguments. In a spatially-symmetric 2-probe system, for example, one can
assume that $\mu= \frac{1}{2} (\mu_1 +\mu_2 )$.

\subsection{Two-probe differential conductance and conductance in more details}
We now apply the above formalism to derive two-terminal conductance formul\ae.
We first consider the case of
two non-superconducting leads (from now on, referred to as NN system) and then
the case where one of the leads is a superconductor (NS system).
$I_i$ denotes the current entering the scatterer from the lead $i\in [1,2]$ and $v_i$ is the bias voltage applied to it.
Current conservation requires that $I_2=-I_1=-I$.

In the NN system the conductance at zero-temperature and zero-voltage can be calculated by
rewriting equation (\ref{a}) in matrix form as
\begin{equation}
\left(
\begin{array}{c}
I\\-I\\
\end{array}
\right)=\left(
\begin{array}{cc}
a_{11}^\uparrow +a_{11}^\downarrow & a_{12}^\uparrow +a_{12}^\downarrow \\
a_{21}^\uparrow +a_{21}^\downarrow & a_{22}^\uparrow +a_{22}^\downarrow \\
\end{array}
\right) \left(
\begin{array}{c}
v_1-v\\v_2-v\\
\end{array}
\right)
\label{II}
\end{equation}
and considering its inverse
\begin{equation}
\left(
\begin{array}{c}
v_1-v\\v_2-v\\
\end{array}
\right)=\frac{1}{d}\left(
\begin{array}{cc}
\tilde{a}_{22} & -\tilde{a}_{12}\\
-\tilde{a}_{21}& \tilde{a}_{11} \\
\end{array}
\right) \left(
\begin{array}{c}
I\\-I\\
\end{array}
\right)
\label{vv}
\end{equation}
where $\tilde{a}_{ij}=a_{ij}^\uparrow + a_{ij}^\downarrow$ and $d=
\tilde{a}_{11} \tilde{a}_{22}- \tilde{a}_{12} \tilde{a}_{21}$ is the
determinant. Since we are interested in the conductance defined as
$G=\frac{I}{v_1-v_2}$, we want to get rid of the condensate chemical potential
$v$. We can achieve this goal by subtracting the second equation in (\ref{vv}) from
the first one, obtaining
\begin{equation}
G= \frac{d}{\tilde{a}_{11} +\tilde{a}_{12} +\tilde{a}_{21} +\tilde{a}_{22}}
\end{equation}
where the matrix elements $\tilde{a}_{ij}$, evaluated at the Fermi energy,
are expressed as follows:
\begin{equation}
\tilde{a}_{11}=\frac{e^2}{h} \left[\left( N^\uparrow -R_0^{\uparrow\uparrow}
-R_0^{\downarrow\uparrow} +R_a^{\uparrow\uparrow} +R_a^{\downarrow\uparrow}
\right) + \left( N^\downarrow- R_0^{\uparrow\downarrow}
-R_0^{\downarrow\downarrow} +R_a^{\uparrow\downarrow}
+R_a^{\downarrow\downarrow} \right)\right] ~,
\end{equation}
\begin{equation}
\tilde{a}_{21}=\frac{e^2}{h} \left[\left( T_a^{\uparrow\uparrow}
+T_a^{\downarrow\uparrow} -T_0^{\uparrow\uparrow} -T_0^{\downarrow\uparrow}
\right) +\left( T_a^{\uparrow\downarrow} +T_a^{\downarrow\downarrow}
-T_0^{\uparrow\downarrow} -T_0^{\downarrow\downarrow} \right)\right] ~,
\end{equation}
\begin{equation}
\tilde{a}_{12}=\frac{e^2}{h} \left[\left( T_a'^{\uparrow\uparrow}
+T_a'^{\downarrow\uparrow} -T_0'^{\uparrow\uparrow} -T_0'^{\downarrow\uparrow}
\right) +\left( T_a'^{\uparrow\downarrow} +T_a'^{\downarrow\downarrow}
-T_0'^{\uparrow\downarrow} -T_0'^{\downarrow\downarrow} \right)\right]
\end{equation}
and
\begin{equation}
\tilde{a}_{22}=\frac{e^2}{h} \left[\left( N'^\uparrow -R_0'^{\uparrow\uparrow}
-R_0'^{\downarrow\uparrow} +R_a'^{\uparrow\uparrow} +R_a'^{\downarrow\uparrow}
\right) + \left( N'^\downarrow- R_0'^{\uparrow\downarrow}
-R_0'^{\downarrow\downarrow} +R_a'^{\uparrow\downarrow}
+R_a'^{\downarrow\downarrow} \right)\right] ~.
\end{equation}
Here $T_0^{\sigma\sigma'}= P_{21}^{+\sigma,+\sigma'}$ ($T_a^{\sigma\sigma'}=
P_{21}^{-\sigma,+\sigma'}$) is the probability for normal (Andreev)
transmission for an injected $\sigma'$-spin quasi-particle from the
left-lead into a $\sigma$-spin quasi-particle (hole) in the
right-lead. For the primed quantities, the quasi-particles are injected from the
right-lead into the left-lead.
Note that $T_0^{\sigma\sigma'}$ and $R_0^{\sigma\sigma'}$,
for $\sigma\neq \sigma'$, and $R_a^{\sigma\sigma}$ and $T_a^{\sigma\sigma}$
are non-zero only if spin-flip scattering is present.
As a check of consistency it is easy to prove that in the absence of
superconductors one recovers the well-known Landauer formula:
\begin{equation}
G=\frac{e^2}{h} \left[ T_0^{\uparrow\uparrow} +T_0^{\downarrow\uparrow}
+T_0^{\uparrow\downarrow} +T_0^{\downarrow\downarrow} \right] ~.
\end{equation}
When superconductors are present, it is interesting to note that their common
condensate chemical potential $\mu$ can be determined imposing the conservation
of the current which yields:
\begin{equation}
v=\frac {(\tilde{a}_{12}+\tilde{a}_{22})v_1+
(\tilde{a}_{22}+\tilde{a}_{21})v_2} {\tilde{a}_{11}+\tilde{a}_{12}+
\tilde{a}_{21}+\tilde{a}_{22}} ~.
\end{equation}

In the NS system, we additionally insist that the superconducting lead and the
superconductors in the scattering region share the same chemical potential:
$v_2=v$. Note that the relation $I_2=-I_1$ allows us to avoid the explicit
calculation of the current in the superconducting lead ($I_2$), which will be a
combination of quasi-particle current and supercurrent.

The conductance at zero-temperature and zero-voltage is also easily derived.
Equation (\ref{II}) becomes trivial since, on the one hand, no transmission of
quasi-particles to the superconductor is allowed at the Fermi energy
($\tilde{a}_{12}=0$) and, on the other, $v_2-v=0$. Therefore (\ref{II})
reduces to $I=\tilde{a}_{11}(v_1-v)$ which gives
\begin{equation}
G=\frac{I}{v_1-v}=\frac{e^2}{h} \left[ \left( N_1^\uparrow-R_0^
{\uparrow\uparrow} -R_0^{\downarrow\uparrow} +R_a^{\uparrow\uparrow}
+R_a^{\downarrow\uparrow} \right) + \left( N_1^\downarrow-
R_0^{\uparrow\downarrow} -R_0^{\downarrow\downarrow} +R_a^{\uparrow\downarrow}
+R_a^{\downarrow\downarrow} \right) \right]
\end{equation}
where the probability coefficients are calculated at the Fermi energy.

It is very instructive to consider a structure composed of a superconducting scatterer
of length $L_{\text{S}}$ attached to two normal leads, one on the left and the other on
the right.
From Eq. (\ref{lambertformula}) one can derive the following 2-probe linear-regime
zero-temperature conductance \cite{lambert93}:
\begin{equation}
G=\frac{2e^2}{h} \left[ T+T_\text{a}+\frac{2(R_\text{a} R_\text{a}'-T_\text{a} T_\text{a}')}
{R_\text{a} +R_\text{a}'+T_\text{a}+T_\text{a}'}\right] ~,
\end{equation}
where $T$ ($T_\text{a}$) is the normal (Andreev) transmission probability for quasi-particles
injected from the left lead and arriving on the right lead.
$R$ ($R_\text{a}$) is the normal (Andreev) reflection probability for quasi-particles
injected from the left lead.
Similarly $T_\text{a}'$ and $R_\text{a}'$ are Andreev scattering probabilities for
quasi-particles injected from the right lead.
For $L_{\text{S}}\gg \xi$ all transmission probabilities tends to zero, since the quasi-particles
penetrate into the superconductor up to a depth of the order of $\xi$, so that the conductance
reduces to the series of two interface resistances, namely $1/2R_\text{a}$ and $1/2R_\text{a}'$
in units of $h/(2e^2)$.
A consequence of this is the fact that the overall conductance of the structure does not depend
on $L_{\text{S}}$, but simply on the microscopic structure of the NS interfaces.
In particular, for $L_{\text{S}}\rightarrow \infty$ although the resistance of the system attains
an asymptotic finite value, the resistance per unit length ({\em resistivity}) vanishes,
as it must be for a superconductor.

According to Eq. (\ref{lambertformula}), the problem of determining the current-voltage
characteristic is reduced to the calculation of scattering amplitudes.
This can be done in different ways.
In simple ballistic systems, for example, by solving the Bogoliubov-de Gennes equation
piecewise in the different homogeneous regions composing the scatterer and imposing proper
matching conditions to the wavefunctions at the boundaries between these regions.
Complex structures such as disordered and diffusive conductors, heterostructures composed
of different materials and multi-terminal systems can be dealt with too.
Scattering amplitudes can be determined numerically, for example, by discretizing the system
in real space within the tight-binding formalism.
For diffusive wires and chaotic quantum dots a different approach can be successfully
employed, namely the random-matrix theory (see, for example, the review papers of
Refs. \cite{stone91,beenakker97}).
According to this theory, from the statistical properties of a class of matrices
with random elements describing a certain physical system it is possible to extract
the properties of the system.
This can be applied to scattering matrices: the first problem consists in determining
their random-matrix probability distribution (statistical ensemble).
The second problem is to find the correlation functions of the transmission eigenvalues
from which the transport properties can be derived.
The conductance, for example, is simply given by
$2e^2/h~\int_0^1 dT T \rho (T)$,
where $\rho (T)$ is the mean eigenvalue density.
A complete solution to this problems has been found for diffusive wires and
quantum dots.
For hybrid systems consisting of a phase-coherent structure connected to superconducting
leads, the random-matrix theory is based on a relationship which links the Andreev
reflection to the transmission eigenvalues of the corresponding normal system
\cite{beenakker92}:
\begin{equation}
\frac{4e^2}{h}\sum_{n=1}^N \frac{T_n^2}{(2-T_n)^2}
\label{nsrelation}
\end{equation}
where $T_n$ is a transmission eigenvalue and $N$ is the number of open channels~\cite{ARED}.
From Eq. (\ref{nsrelation}) it is clear that hybrid systems can be studied making
use of the results developed for normal conductors.

\subsection{Quasi-classical Green's function approach}
\label{Green}

In this section we briefly underline another very important technique for dealing 
with hybrid systems.
This approach applies when the characteristic length scale of the problem are
large if compared with Fermi wavelength.
It proves to be particularly useful for dirty structures where a
sufficient amount of non-magnetic impurities is present so to make the motion of
electrons isotropic.

The quasi-classical theory of superconductivity is developed through quantum field
theory methods and it is formulated in terms of
Green's functions in the particle-hole space (known as {\em Nambu space}).
Such Green's functions are denoted by a {\em hat} and, in addition to standard Green's
functions $G$, they contain anomalous components $F$ which describe the superconducting
correlations:
\begin{equation} \hat{G}(E,\mathbf{r}_1,\mathbf{r}_2)=
\left( \begin{array}{cc} G(E,\mathbf{r}_1,\mathbf{r}_2) & F(E,\mathbf{r}_1,\mathbf{r}_2) \\
F^{\dagger}(E,\mathbf{r}_1,\mathbf{r}_2) & G^{\dagger}(E,\mathbf{r}_1,\mathbf{r}_2)
\end{array}\right) ~.
\end{equation}
The starting point is the Gorkov equation \cite{gorkov58} for the Green's function
of a bulk superconductor which was derived directly from the BCS Hamiltonian.
The quasi-classical approximation consists in averaging the Green's function $\hat{G}$
over the rapid oscillations in the relative space coordinates
$\mbox{\boldmath $\rho$}=\mathbf{r}_1-\mathbf{r}_2$ and over impurities.
As a results one obtains the quasi-classical Green's function
$\hat{g}(E,\mathbf{r},\mathbf{n})$ which depends on the energy $E$, on the center-of-mass
coordinate $\mathbf{r}=(\mathbf{r}_1+\mathbf{r}_2)/2$ and the versor of
the relative momentum $\mathbf{n}$ associated to the relative coordinate $\mbox{\boldmath $\rho$}$.
The Gorkov equation reduces to the following (Eilenberger) equation for $\hat{g}$:
\begin{equation}
\mathbf{v}_{\text{F}}\partial_{\mathbf{r}}\hat{g}+\left[
-iE\hat{\sigma}_3 +\hat{\Delta}+\frac{\langle \hat{g} \rangle}{2\tau} , \hat{g}
\right]=0~,
\label{eileq}
\end{equation}
with
\begin{equation}
\hat{\Delta}=
\left( \begin{array}{cc} 0&
\Delta \\\Delta^* &0
\end{array}\right)
\end{equation}
derived in Refs. \cite{eilenberger68,larkinovchinikov69}.
Here $\mathbf{v}_{\text{F}}$ is the Fermi velocity, $\Delta$ is the pair potential
and $\tau$ is the elastic scattering time.
Furthermore the square brackets represent the commutator, while the angular brackets
$\langle \cdots\rangle$ denote an averaging over the direction $\mathbf{n}$.
Note that Eq. (\ref{eileq}) determines $\hat{g}$ up to a multiplicative constant
and the following normalization condition must be applied: $\hat{g}^2=\hat{1}$.

A further simplification can be realized in the presence of an isotropic impurity
scattering potential which makes the motion of electrons diffusive (dirty limit).
In this case $\hat{g}$ can be expanded in spherical harmonics and Eq. (\ref{eileq})
reduces to the Usadel equation \cite{usadel70} for the isotropic function
$\hat{g}(E,\mathbf{r})=\langle \hat{g}(E,\mathbf{r},\mathbf{n}) \rangle$:
\begin{equation}
D\partial_\mathbf{r}\left( \hat{g}\partial_\mathbf{r}\hat{g} \right)+
\left[ iE\hat{\sigma}_3-\hat{\Delta},\hat{g} \right]=0
\label{usadeleq}
\end{equation}
where $D=v_{\text{F}}^2\tau/3$ is the diffusion constant and $v_{\text{F}}$ is the magnitude of the Fermi velocity.
It is worthwhile to stress that both Eq. (\ref{eileq}) and (\ref{usadeleq}) must
be supplemented with a self-consistent condition for the pair potential $\Delta$.
The electrical current can then be calculated once the Green's function $\hat{g}$ is
determined.

It should be noted that the quasi-classical approximation does not allow to take into
account nonuniformities that occur on the Fermi wavelength scale,
such as boundaries, barriers and interfaces with other materials.
It was shown, however, that this problem can be circumvented by
applying proper boundary conditions to the quasi-classical Green's function.
For the Eilenberger equation they were derived in Ref. \cite{zaitsev84} and for the
Usadel equation in Ref. \cite{kupriyanov88}.
It is interesting to note that such boundary conditions are obtained by
making use of the connection between scattering amplitudes and Green's functions.
For a more detailed treatment of the quasi-classical theory we refer the reader to a
number of review papers on the subject (for the most recent see, for example, Refs.
\cite{rammersmith,raimondi,shon,beenakker}).

An important progress in the quasi-classical theory was put forward by Nazarov
who formulated it, within the dirty limit, in terms of a
circuit theory \cite{nazarov94,nazarov95,nazarov99}.
This constitutes a generalization of the classical Ohm's law to hybrid coherent
nanostructures. These can be viewed as
coherent networks consisting of {\em nodes} linked by {\em connectors}.
The starting point consists in observing that when the second term in the Usadel
equation (\ref{usadeleq}) can be neglected, the same equation can be written in the
form of a conservation law for a matrix current $\check{\mathbf{j}}(\mathbf{r})$
\cite{Nambu-Keldysh}:
\begin{equation}
\partial_{\mathbf{r}}\check{\mathbf{j}}(\mathbf{r})=0;~~~~
\check{\mathbf{j}}(\mathbf{r})=\sigma \check{G}\partial_{\mathbf{r}} \check{G}
\label{conservation}
\end{equation}
where $\sigma$ is the normal state conductivity proportional to the diffusion constant $D$.
The second equation in (\ref{conservation}) resembles the local Ohm's law in a
normal metal structure, while the first expresses the conservation of the current.
The problem is finally defined by imposing additional boundary
conditions, which sets the bias voltage established across the structure and describe
interfaces in the conductor.
By solving those equations the voltage distribution over the conductor and hence
the local current density and total current can be determined.
The ordinary circuit theory is formulated by discretizing
Eqs. (\ref{conservation}) plus boundary conditions through a
finite element approximation so that the structure can be separated into resistive
elements pairwise connected.
Current conservation in each node and current-voltage relations for each element
set the voltage in each node allowing the determination of the total current.

The extended circuit theory cannot be derived in such a simple way because of the
matrix form of the equations and the fact that one deals with matrix currents and
Green's functions.
However the situation can be simplified due to the symmetry properties of the Green's
function at zero energy so that Eqs. (\ref{conservation}) can be separated into two
parts.
The first one determines the equilibrium spectral local properties and the second one the
non-equilibrium chemical potentials in each node which determines the electron propagation.
These equations allows to formulate a circuit theory which can be conveniently condensed
into a set of rules (see Ref. \cite{nazarov94}) for determining currents and potential
differences of a coherent network of nodes and connectors.
It is worth mentioning that: i) nodes are required to contain enough disorder to
prevent ballistic transmission between the two connectors attached to it.
Furthermore the size of a node must be smaller than the coherence length.
ii) connectors can be of three types, namely diffusive conductors, tunnel junctions or
ballistic constrictions. The net effect of superconductivity is to induce a renormalization
of the conductivity in the last two types of connectors.

It is important to recall that the above circuit theory is valid when the second term in the
Usadel equation can be neglected, i.e. for small temperatures and voltages
\cite{nazarov94}: $k_BT,eV\ll \Delta, \hbar D/L^2$, $L$ being the system size.
The limitations described above are quite restrictive, namely the voltage and temperature
dependence of the transport properties cannot be accessed.
However, a circuit theory can still be formulated by discretizing
the Usadel equation (\ref{usadeleq}) comprised of the second term.
This can be done by discretizing the space into a set of nodes $x_i$ such that
the Green's function $\check{G}(x_i)\equiv\check{G}_i$ in neighboring points are close to each
other.
A resistor is associated with each connection.
The Eq. (\ref{usadeleq}) can then be viewed as a conservation law in each node $i$
($\sum_k\check{I}_{ik}+\check{I}_{\text{leakage},i}=0$)
where now the total current is composed of $\check{I}_{ik}$ (which corresponds to
the second expression in Eq. (\ref{conservation})) and an additional {\em leakage}
contribution $\check{I}_{\text{leakage},i}\propto -ie^2[\check{G}_i,\check{H}]$
associated to the second term of the Usadel equation (\ref{usadeleq}).
The leakage current describes two processes, the first one, proportional to the energy
$\epsilon$, is associated to the decoherence of electrons and holes due to wave-vector
mismatch at finite energies (``leakage'' of coherence).
The second one, proportional to $\Delta$, is responsible for the conversion between
quasi-particles and Cooper pairs (``leakage of quasi-particles'').
In order to obtain an accurate agreement with the full theory the distance between the
nodes has to be smaller than the coherence length $\xi=\sqrt{D \text{Max}(\Delta,\epsilon)}$.
It is important to notice that ballistic point contacts can be considered as a particular
kind of connectors, but they need a specific treatment.
To conclude, this new circuit theory can be formalized into a set of rules given in
Ref. \cite{nazarov99}.

\subsection{Panorama on results in NS systems}
\label{Results}

Superconducting hybrid nanostructures possess a rich physics related, in particular,
to the interplay between coherent transport of electrons and Andreev reflection.
Physical systems investigated so far comprise NS junctions, mesoscopic SNS Josephson
junctions, structures containing several superconducting islands and, more recently,
ferromagnet/superconductor (FS) junctions, where the interplay between spin-dependent
transport and superconductivity can be investigated.
The vast knowledge available in semiconductor technology has been also exploited by
replacing in the above
systems normal-metals with semiconductors (Sm). 
The possibility of using unconventional superconducting materials, such as d-wave
superconductors and, more generally, high-T$_\text{c}$
superconductors, has been explored too.
As a result, a variety of effects and phenomena have been reported so far, the
most important of whose we shall review in the present section making use of
the various theoretical approaches outlined in the previous sections.
It is worth mentioning that these allow one to calculate, in addition to the electrical
current, all transport properties of a given system such as thermal current,
current noise and the whole full counting statistics of electronic transport.

We shall start considering NS junctions, which present different properties depending
on whether N is diffusive or ballistic and on the transparency of the barrier at the interface.
Perhaps the first effect that received much attention is the
zero bias anomaly (ZBA), which
refers to a conductance peak observed at zero bias voltage in a low-transparency NS point contact
\cite{kastalsky91}.
This contrasts with the BTK result, where the conductance presents a minimum at zero bias,
is particularly striking since the
height of the peak is of the order of the normal state conductance.
ZBA can be explained as an interplay between Andreev reflection at the interface with
the superconductor and disorder scattering in the N
side of the junction due to the presence of impurities.
When the elastic mean free path in the N side of the junction is smaller
than the junction size, electrons have the chance to be scattered
back to the S interface many times, finally undergoing an Andreev reflection.
The net result is that Andreev reflection processes occur with a much larger
probability with respect to the case of absence of impurities.
In other words the low bias conductance is determined by complex interference effects
which produce an enhancement by several orders of magnitude.
Such an effect dies away for energies larger than the Thouless energy which sets the
scale for particle-hole dephasing. 
Although ZBA was first understood within the quasi-classical approach, it was
then confirmed using scattering methods and also in the tunneling Hamiltonian
approach.
This effect is also known as {\em reflectionless tunneling}, since it can be
explained in terms of the disorder-induced opening of tunneling channels.

Non-monotonic behaviors of the conductance as a function of voltage and temperature
in NS junctions with clean interfaces were also reported.
When a diffusive wire is connected to a superconducting reservoir through a highly
transparent contact the zero-temperature, zero-voltage average conductance equals
the normal-state conductance \cite{artemenko79,beenakker92}.
Even though this can be proved rigorously, intuitively it is the consequence of two facts,
namely that the
Andreev reflection effectively doubles the length of the normal metal conductor and that
the conductance in a NS point
contact is doubled with respect to the normal-state one.
However it was found \cite{artemenko79,nazarov96,golubov97,yip95,volkov96,lesovik97}
that a conductance peak appears either for voltages or
temperatures of the order of the Thouless energy $\hbar D/L^2$.
This phenomenon, known as {\em reentrance effect}, was experimentally proved
in Refs. \cite{vanwees94,courtois96,charlat96}.

Another interesting phenomenon is the Andreev interferometry which is realized in
a hybrid structure containing at least two superconducting islands.
This is based on the fact that when a quasi-particle Andreev reflects from a NS
interface, the phase of the outgoing excitation is shifted by the phase of the
superconducting order parameter.
In the presence of two superconducting islands with order parameter phases $\phi_1$ and
$\phi_2$, the transport properties are oscillatory functions of the difference
$\phi_1-\phi_2$.
The conductance of individual samples was found to be $2\pi$-periodic in the phase
difference.

During the last few years the interplay between Andreev scattering
and the ferromagnetic order has been addressed in numerous
studies of electronic transport properties in nanostructures
containing both ferromagnets and superconductors.
The interest comes from the fact that the electron and the hole involved
in the Andreev process must belong to opposite spin band.
This produces a suppression of the current flowing through a FS interface,
which depends on the value of the exchange field characteristic of the ferromagnet.
This phenomenon has been used for estimating the spin-polarization of different
kind of ferromagnets \cite{soulen98,upadhyay98}.

\section{Electronic Cooling in Superconducting Nanostructures}
\label{cooling}

\subsection{SIN refrigerators}
\label{sec:SINIS}

\begin{figure}[t!]
\begin{center}
\includegraphics[width=\columnwidth,clip]{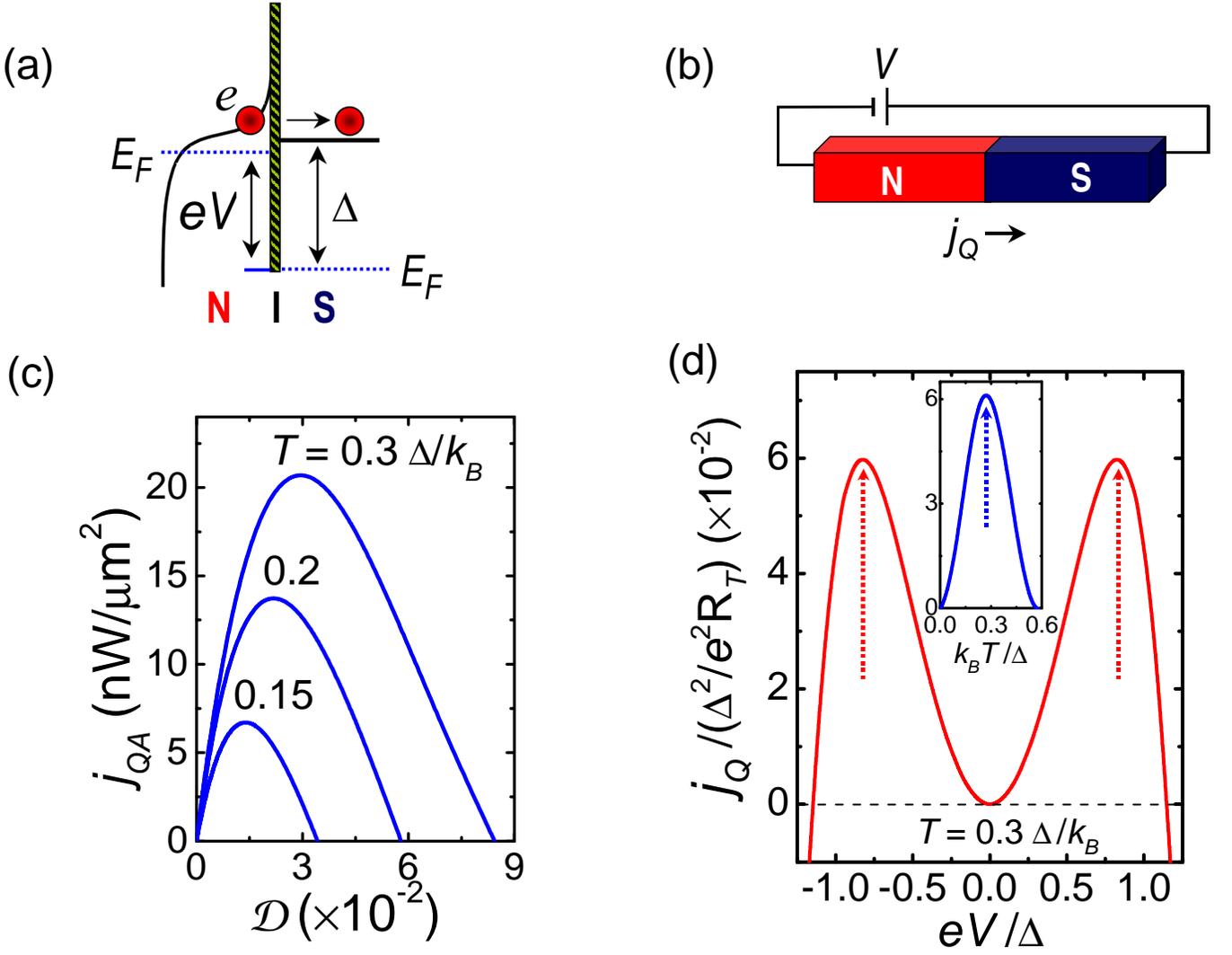}
\end{center}
\caption{(a) Energy band diagram of a biased superconducting tunnel junction. More energetic electrons ($e$) can more easily tunnel into the superconductor. The electron gas in the N electrode is thus \textit{cooled}. (b) Schematic representation of a voltage biased SN junction. Upon biasing the structure, a heat current $j_{Q}$ flows across the system. (c) Maximum specific cooling power $j_{QA}$ versus interface coefficient of transmission $\mathcal{D}$ at three different temperatures. (d) Dimensionless cooling power $j_{Q}$ of a SIN contact versus bias voltage $V$ at $T=0.3\,\Delta/k_B$. The inset shows $j_{Q}$ calculated at the optimal bias voltage versus temperature.}
\label{fig:coolprinciple}
\end{figure}

An important research topic is represented by heat transport in superconducting nanostructures.
In particular, heat transport through SN interfaces can be successfully applied to microcooling \cite{nahum:3123,leivo:1996}. In order to gain some insight into this problem it is instructive to consider the simplified energy band diagram of a NIS \textit{tunnel} junction (I is an insulating barrier) biased at voltage $V$, as schematically depicted in Fig. \ref{fig:coolprinciple}(a).
The physical mechanism underlying electronic cooling is rather simple. 
As discussed in section \ref{solveBdG},
when a normal metal is brought in contact with a superconductor, quasiparticle transport is effective only at energies larger than the superconducting gap $\Delta$. 
As a consequence of this "selective" tunneling of hot particles, the electron distribution function in the N region becomes sharper, i.e., the  electron effective temperature in N is \textit{lowered} (even in the regime when electrons are thermally decoupled from the lattice): the SIN junction thus behaves as an electron cooler.
This situation can be experimentally accessed when transport is dominated by quasiparticle dynamics, i.e., in NIS \textit{tunnel} junctions.

To highlight the impact of junction transmissivity in governing heat flux (cooling power) $j_Q$ \cite{bardas:12873} through a voltage biased SN contact (see Fig. \ref{fig:coolprinciple}(b)) let us consider Fig. \ref{fig:coolprinciple}(c), where the maximum cooling-power surface density  $j_{QA}$ versus interface transparency $\mathcal{D}$  is calculated at different temperatures.
$j_{QA}$ is a non monotonic function of interface transmissivity, going to zero both at low and high values of $\mathcal{D}$. In the low transparency regime it turns out to be linear in $\mathcal{D}$ \cite{bardas:12873},
i.e., the electron transport is dominated by single particle tunneling. 
However, by increasing interface transmissivity, two-electron Andreev  tunneling \cite{andreev64} begins to dominate quasiparticle transfer across the junction leading to enhanced conductivity but suppressing the heat flow extraction from the N metal. 
$j_{QA}$ is maximized at an optimal $\mathcal{D}$ value, which in addition depends on the temperature. 
In real SIN structures for cooling applications (i.e., those exploiting \textit{tunnel} junctions), $\mathcal{D}$ is typically in the range $10^{-6}\div 10^{-4}$ \cite{nahum:3123,leivo:1996} corresponding to junction specific resistances (i.e., the product of the junction normal state resistance and the contact area) from tens to several thousands $\Omega\,\mu$m$^2$. This limits  the achievable $j_{QA}$ to values much lower than those expected from theory, in particular  to some pW$/\mu$$m^{2}$. Barrier optimization seems to be still nowadays a challenging task.

In the tunnel limit (that is particularly relevant in the experiments), the cooling power is given by  
\begin{equation}
\label{eq:jQ}
j_Q(V)=\frac{1}{e^2R_T}\int_{-\infty}^{\infty}dE \tilde{E}N_S(E)[f_N(\tilde{E})-f_S(E)],
\end{equation} 
where $\tilde{E}=E-eV$, $f_{S(N)}(E)=f_0(E,T_{S(e)})$ is the Fermi-Dirac distribution of quasiparticles at temperature $T_S$ in the superconducting reservoir (and of electrons at temperature $T_e$ in the N electrode),
$R_T$ is the normal-state resistance of the junction and $N_S(E)=|\textrm{Re}[E/\sqrt{E^2-\Delta ^2}]|$ is the BCS normalized density of states of the superconductor.
Figure \ref{fig:coolprinciple}(d) shows  $j_Q$ calculated from Eq. (\ref{eq:jQ}) versus bias voltage at  $T=T_e=T_S=0.3~\Delta/k_B$.
The cooling power is a symmetric function of the bias voltage $V$, i.e., $j_Q(V,T)=j_Q(-V,T)$, and, at this temperature, it is maximized around $V\approx \pm 0.8~\Delta/e$, as indicated by red-dashed arrows in the figure.  
Note that when $j_Q$ is positive, it implies heat removal from the N electrode.
Furthermore, for each temperature, there is an optimized voltage that maximizes $j_Q$. 
The inset of Fig. \ref{fig:coolprinciple}(d) displays the heat current versus temperature calculated at each optimal bias voltage. The cooling power is maximized at $T\approx 0.3~\Delta/k_B=T_{opt}$ (indicated by the blue-dashed arrow) where it reaches $j_Q\simeq 6\times 10^{-2} \Delta^{2}/e^2R_T$, decreasing both at lower and higher temperatures.

The final $T_e$ is determined from the balance among several factors that tend to drive power into the electron  system. The majority of SIN microrefrigerators realized so far operate in a regime  where strong inelastic electron-electron interaction  drives the system to (\textit{quasi})equilibrium, where the electron system can be  described by a Fermi function at a  temperature $T_e$ that differs, in general, from that of the lattice ($T_{ph}$). At the  temperatures of interest (i.e., typically below 1 K ), the predominant contribution comes from electron-phonon scattering that exchanges energy between electrons and the lattice phonons. This rate of exchanged energy is given by \cite{kautz:2386,wellstood:5942}
\begin{equation}
\label{eq:ephint}
j_Q^{e-ph}=\Sigma \mathcal{V}(T_{ph}^5-T_e^5),
\end{equation}  
where $\Sigma$ is a material-dependent parameter of the order $10^{-9}$ WK$^{-5}\mu$m$^{-3}$ and $\mathcal{V}$ is the volume on the normal electrode. 
The refrigerator cooling power ($j_Q^{refr}$) can be defined as the maximum power load that the SIN device can sustain while keeping the N region at temperature $T_e$, and can be generally expressed as 
\begin{equation}
\label{eq:ephbaleq}
j_Q^{refr}=j_{Q}(V,T_e,T_{ph})-j_Q^{e-ph}(T_e,T_{ph}).
\end{equation}  
The minimum achieved temperature $T_e$ thus fulfills the condition $j_Q^{refr}=0$.

The first observation of heat extraction from a normal metal dates back to 1994 \cite{nahum:3123}, where cooling of electrons  in Cu below the lattice temperature was demonstrated using an Al/Al$_2$O$_3$/Cu tunnel junction. An important improvement was made two years later, still in the Al/Al$_2$O$_3$/Cu system, by Leivo \textit{et al.} \cite{leivo:1996}, which recognized that using two SIN junctions in series arranged in a symmetric configuration (i.e., in a  SINIS fashion) would have led to a much stronger cooling effect. In this configuration a reduction of the electron temperature from $300$ mK to about $100$ mK (i.e., close to the achievable optimum) was obtained. Later on, several other experimental evidences of electron cooling in SINIS metallic structures were reported \cite{leoni:3877,fisher:2705,clark:625,arutyunov:326,leoni:3572,tarasov:714,pekola:2782,pekola:485,vystavkin:598,fisher:561,leivo:227,pekola:056804,luukanen:281}. In these experiments SIN junctions are used  to alter the electron temperature in the N region as well as to measure it. In order to measure the temperature, the N region is normally connected to other SIN contacts (i.e., "probing" junctions) that act as thermometers, suitably calibrated by varying the bath temperature of the cryostat.

\subsection{SF refrigerators}
\label{sec:FS}

\begin{figure}[t!]
\begin{center}
\includegraphics[width=\columnwidth,clip]{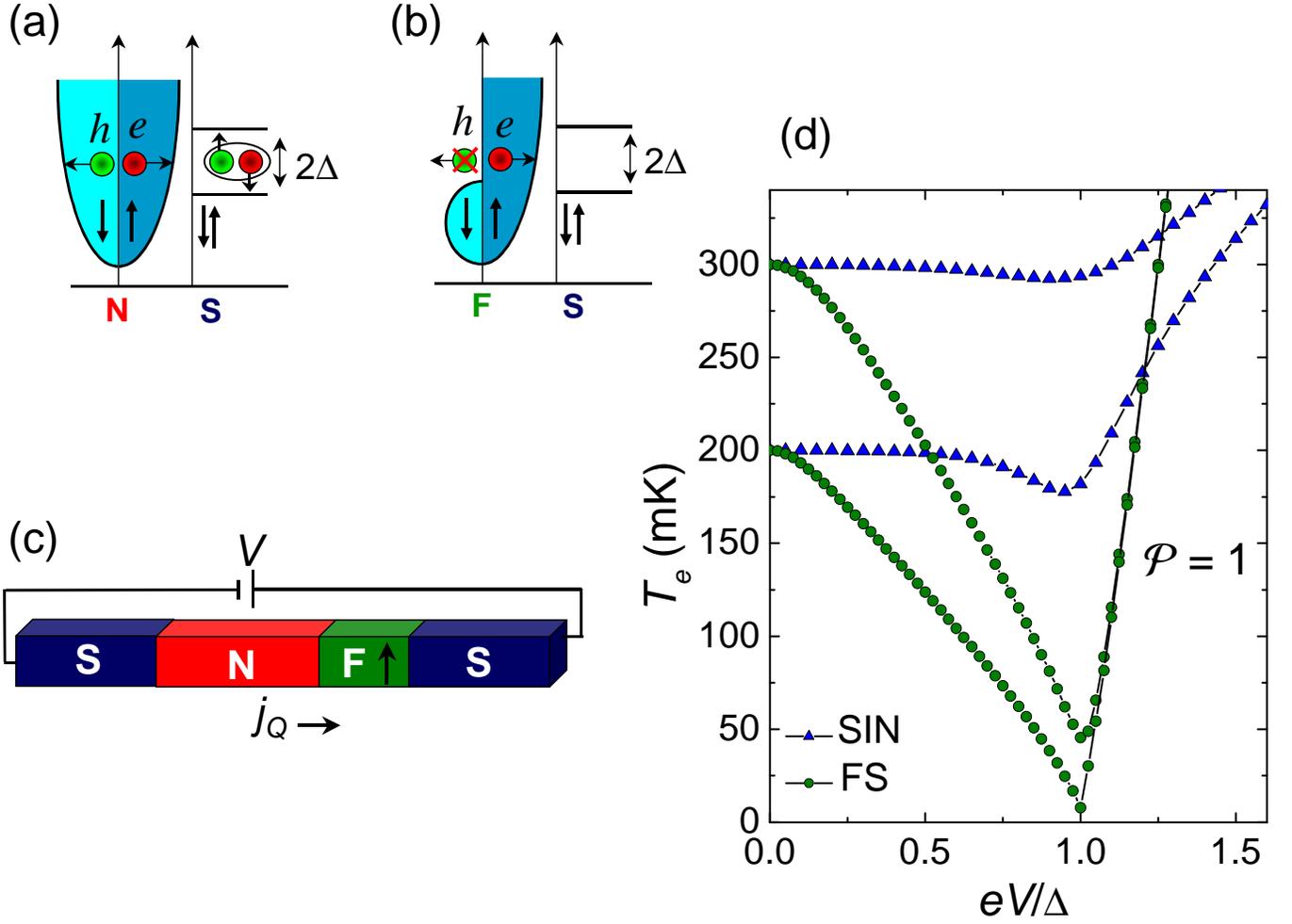}
\end{center}
\caption{(a) Schematic description of Andreev reflection at a SN contact. (b) Schematic representation of the principle of operation of a SFN electron cooler. For $\mathcal P=1$, Andreev reflection is hindered by the absence of available states for reflected holes ($h$). This subgap electron-transport suppression mechanism allows the operation of the microrefrigerator in the presence of efficient carrier transfer to S. (c) Sketch of the SFN electron cooler. The FN contact is supposed to be a highly transmissive electric contact.
(d) Electron temperature $T_e$ vs bias voltage $V$ for a half-metallic ($\mathcal P=1$) SF and SIN microcoolers for two starting bath temperatures $T_{ph}$ at $V=0$.
Adapted from Ref. \cite{giazotto:3784} and Ref. \cite{giazotto:japan}.}
\label{fig:fsjunction}
\end{figure}

As discussed in Sec. \ref{sec:SINIS},
decreasing the contact resistance $R_T$ is not a viable route to enhance heat current in SIN junctions. 
As a matter of fact, the increase of the Andreev  reflection contribution to the electric transport \cite{andreev64} (see Fig. \ref{fig:fsjunction}(a))  strongly affects and limits thermal current through the system. 
A promising  scheme to  surmount this problem is to exploit, instead of an insulating  barrier, a thin ferromagnetic (F) layer in good electric contact with S  \cite{giazotto:3784}. 
As already discussed in section \ref{BdGFS}, depending on the degree of spin polarization $\mathcal{P}$ of the F layer, strong suppression of the Andreev current may occur at the SF interface \cite{dejong95}. In the limit of large $\mathcal{P}$, the subgap  current is thus strongly suppressed while efficient hot-carrier transfer leads to a considerable heat current across the system. 
The authors of Ref. \cite{giazotto:3784} focused on a ballistic three-dimensional SF junction (as shown in Fig. \ref{fig:fsjunction}(c)) where the F region is described by the usual Stoner model \cite{dejong95}; furthermore, the analysis was carried out along the lines of a voltage biased SN junction \cite{bardas:12873,BTK82} but generalizing heat transport to a spin-dependent superconducting structure. In the following we shall give the main results of such analysis.  

Figure \ref{fig:fsjunction}(d) shows $T_e$ as a function of bias  voltage for 
two starting lattice temperatures (i.e., the system temperatures at $V=0$) $T_{ph}=200$ mK and $T_{ph}=300$ mK.
The calculation 
was performed for half-metallic (${\mathcal P} =1$) SF (filled-green dots) and for SIN (filled-blues triangles)
microcoolers, assuming Al as superconductor 
($\Delta_{Al}= 180 \,\,\mu eV$).
In particular, here it was assumed a junction area $\mathcal A =0.1\,\mu$m$^2$, an active 
    volume of the FN side 
    ${\mathcal V} =0.5$ $\mu$m$^3$, $\Sigma =2\times 10^{-9}$ WK
    $^{-5}$$\mu$m$^{-3}$ (typical of Cu), 
    and a SF normal-state resistance $R_{n}^{FS}=28$ m$\Omega$ (assuming a Fermi energy ${\mathcal E}_{F}=5\,\,eV$), 
and for the 
SIN-junction specific contact resistance a value  $0.4$ k$\Omega$\,$\mu$m$^2$, corresponding to high-quality Cu/Al$_2$O$_3$/Al junctions~\cite{nahum:3123,leivo:1996}.
Figure \ref{fig:fsjunction}(d) shows the remarkable  $T_e$ reduction provided by  the 
FS cooler with respect to the NIS cooler. 
Even starting from 200 mK, for which the tunnel junction provides its largest cooling effect (a temperature reduction of about 10\% at $eV\simeq \Delta$), the SF cooler
yields $T_e$ of the order of 
10 mK (a temperature reduction of about 95\%). This marked difference stems from the high contact resistance 
of the SIN junction that strongly affects its performance and must be compared
to specific contact resistances as low as $10^{-3}\,\,\Omega \mu$m$^2$ that are currently achieved in highly transmissive SF junctions~\cite{upadhyay98}.

\begin{figure}[t!]
\begin{center}
\includegraphics[width=\columnwidth,clip]{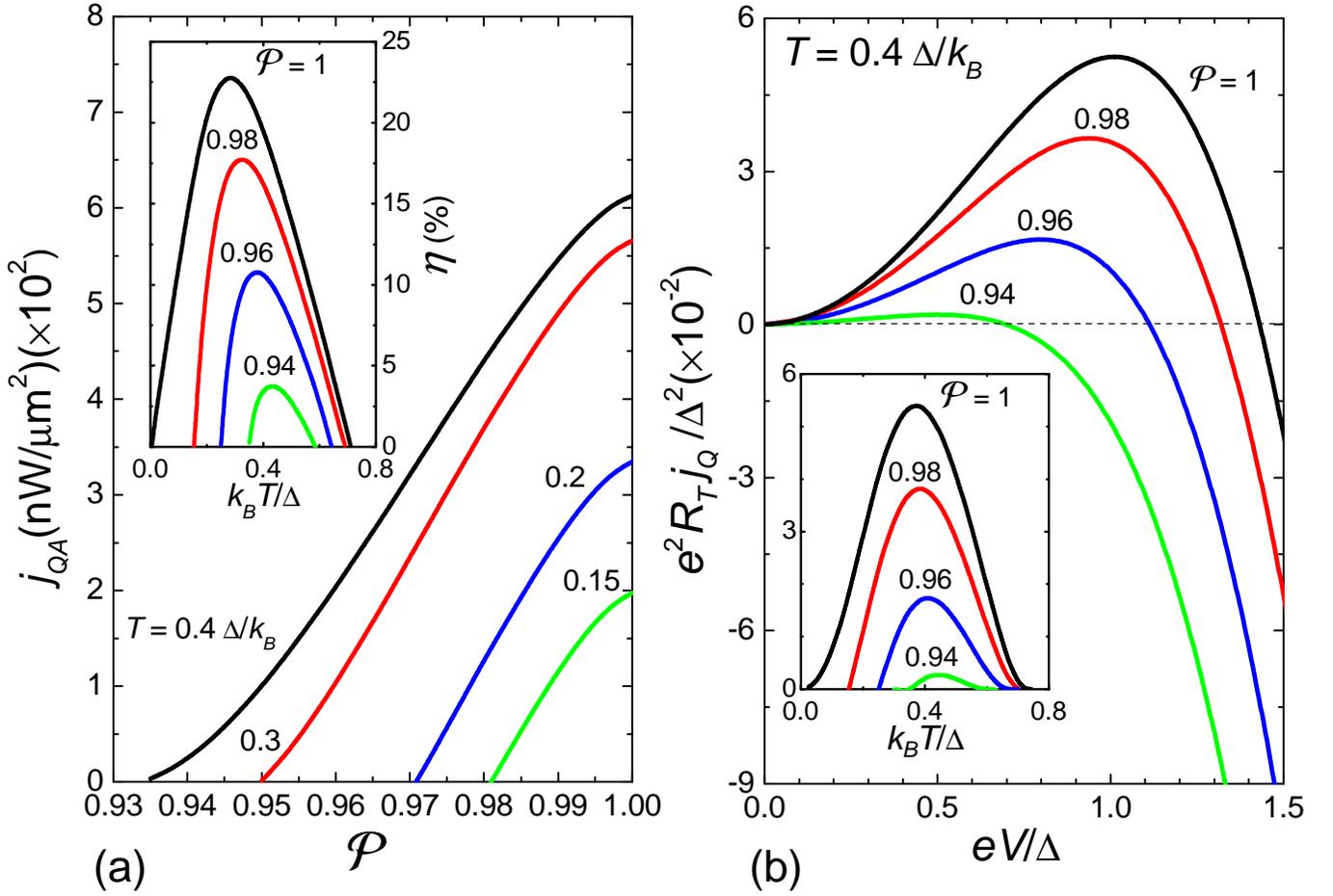}
\end{center}
\caption{(a) Maximum cooling-power surface density $j_{QA}$ versus  spin polarization $\mathcal P$ for various temperatures. The inset shows the coefficient of performance $\eta$ calculated at the optimal bias voltage versus temperature for some $\mathcal P$ values. (b) Dimensionless cooling power $j_{Q}$ vs bias voltage at $T=0.4\,\Delta/k_B$ for some $\mathcal P$ values. The inset shows the heat current calculated at the optimal bias voltage against temperature for several $\mathcal P$ values.
Adapted from Ref. \cite{giazotto:3784} and Ref. \cite{giazotto:japan}.}
\label{fig:fsjunction2}
\end{figure}

The junction specific cooling power is shown in Fig. \ref{fig:fsjunction2}(a), where $j_{QA}$ (evaluated at each optimal bias voltage) is plotted versus $\mathcal{P}$ at different bath temperatures. Notably, for a half-metallic ferromagnet at $T=0.4~\Delta/k_B$ cooling power densities  as high as $600$ nW/$\mu$m$^2$ can be achieved, i.e., about a factor of 30 larger than those achievable in SIN junctions at the optimized interface transmissivity (i.e., $\mathcal{D}\simeq 3\times 10^{-2}$ at $T\simeq 0.3\Delta/k_B$, see Fig. \ref{fig:coolprinciple}(c)).
The impact of partial spin polarization ($\mathcal{P}<1$) in the F region is shown in Fig. \ref{fig:fsjunction2}(b) where the normalized heat current $j_Q$ versus bias voltage across the junction is plotted for some $\mathcal{P}$ values at $T=0.4~\Delta/k_B$. For each $\mathcal{P}$ value there exist an optimum bias voltage which maximizes $j_Q$. In the limit of a half-metal ferromagnet (i.e., $\mathcal{P}=1$) \cite{mazin:1427,coey:988} this value is around $V_{opt}\simeq \Delta/e$. In the inset of Fig. \ref{fig:fsjunction2} (b) is displayed $j_Q$ calculated at each optimized bias voltage against temperature for various $\mathcal{P}$ values. The heat current is maximized around $T=T_{opt}\approx0.4~\Delta/k_B$, rapidly decreasing both at higher and lower temperatures
The inset of Fig. \ref{fig:fsjunction2}(a) shows the junction coefficient of performance ($\eta$) calculated at the optimal bias voltage   versus temperature  for various $\mathcal{P}$ values. Notably, for $\mathcal{P}=1$, $\eta$ reaches $\sim 23\%$ around $T=0.3~\Delta/k_B$ and exceeds $10\%$ for $\mathcal{P}=0.96$. In addition, in light of a possible exploitation of this structure with a N region (i.e., a SFN cooler) it turned out that cooling effects comparable to the SF case can be achieved  with F-layer thickness of  a few nm (i.e., of the order of the magnetic lenght).

The above given results point out the necessity of strongly spin-polarized ferromagnets for a proper operation of the SF cooler. 
Among the predicted half-metallic candidates it is possible to indicate CrO$_2$ \cite{kamper:2788,brener:16582,schwartz:L211}, for which values of $\mathcal{P}$ in the range $85\div 100\%$  have been reported \cite{ji:5585,parker:196601,dedkov:4181,coey:3815}, (Co$_{1-x}$Fe$_{x})$S$_2$ \cite{mazin:3000}, NiMnSb \cite{degroot:2024}, Sr$_2$FeMoO$_6$ \cite{kobayashi:677} and NiMnV$_2$ \cite{weht:11006}. So far no experimental realizations of SF structures for cooling applications were reported. The main technological problems seem, still nowadays, to avoid degradation of  ferromagnetic properties at the SF interface and a proper integration with suitable superconductors.

\section{Coulomb blockade in hybrid structures}
\label{blockade}

\subsection{Coulomb blockade and superconductivity}
The astonishing progress in modern technology has made it possible to fabricate in 
a controlled way tunnel junctions with capacitances of the order of $C = 10^{-15} $F
and below. In this case even the charging energy associated with a single-electron 
accumulation at the junction, $E_{\rm C} \equiv e^2/2C$, can be of the order of 
several Kelvins. This implies that the properties in the sub-Kelvin regime are strongly 
affected by the presence of charging effects. Indeed charging energy allows the control 
of the electron number of small islands with precision $e$ or $2e$ in the case of 
superconducting devices. Adding one electron to a small superconducting island puts
it into an excited state with an energy exceeding the gap. Only when a
second electron is added, can both recombine to form a Cooper pair.
This leads to {\em parity effects}, i.e. to different properties of the 
superconducting systems depending on the number of electrons being even or odd.

The relevant mechanisms for charge transport in superconducting single-electron devices 
are quasi-particle tunneling, two-particles tunneling and coherent tunneling of 
Cooper pairs. All of them are strongly affected by charging. Tunneling of quasi-particles
is very similar to the case of normal metals but taking into account the modified 
BCS density of states. Two-particle tunneling is a higher order process related to 
the Andreev reflection discussed previously. Both two-particle tunneling and coherent 
oscillations of Cooper pairs dominate transport at very low voltages.
Moreover, the interplay of charging and coherent tunneling has a deep significance because it 
leads to the possibility of observing macroscopic quantum dynamics in these systems.
The charge and the phase difference in a Josephson junction, although being macroscopic 
degrees of freedom, are quantum mechanical conjugate variables. Therefore the eigenstates 
are in general superpositions of different charge states. In addition the combination of 
coherent Cooper pair tunneling and quasi-particle tunneling, leads to
a variety of  structures in the $I$-$V$ characteristic.

There are numerous examples of systems where single-electron properties and 
superconductivity have been studied theoretically and realized experimentally. 
A prototype example which we will analyze is the NSN-SET transistor
where a superconducting island is connected to two normal conducting leads
by means of tunnel junctions. This example allows to discuss both quasi-particle 
and two-particles tunneling. We then discuss some basic properties of the 
coherent oscillations of Cooper pairs which will be important for the discussion
on the implementation of quantum computers by means of superconducting nanocircuits.

\begin{figure}
\centerline{\psfig{figure=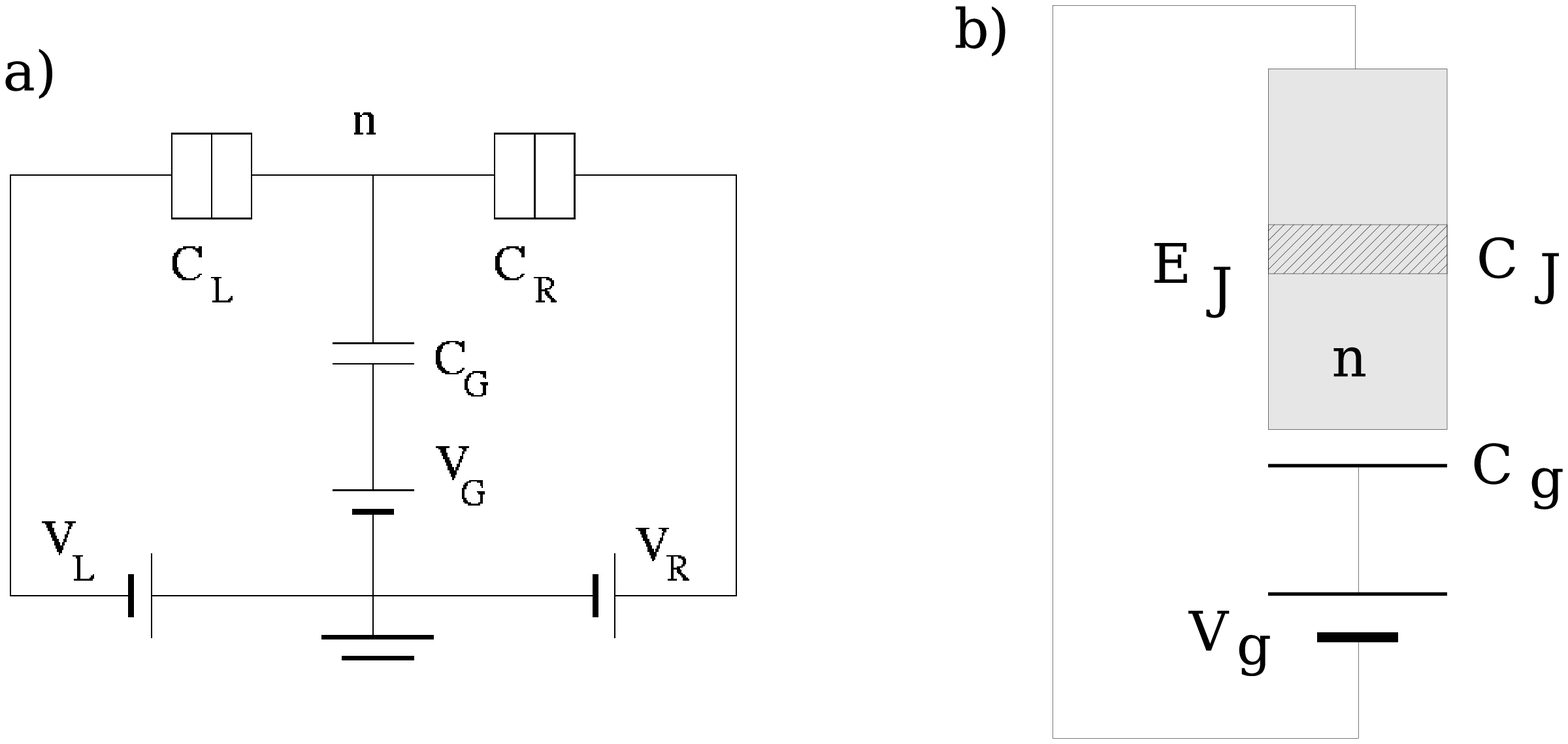,height=7cm,width=14.0cm}}
\caption{a) The SET transistor. b) The Cooper-pair box.}
\label{transistor}
\end{figure}

As prototype examples to discuss charging effects in superconducting nanostructures 
we consider the single-electron (SET) transistor shown in Fig.~\ref{transistor}a and
the single-electron box of Fig.~\ref{transistor}b.
Here we concentrate on the transistor, while
the box will be analyzed in the section 
devoted to quantum computation.
The charging energy of the SET transistor depends on the electron number 
in the central island and on the applied voltages.  The central island is coupled via 
two tunnel junctions to a transport voltage source, $V=V_{\rm L}-V_{\rm R}$, 
so that a current can flow. The island is, furthermore, coupled capacitively 
to a gate voltage $V_{\rm G}$. The charging energy of the system depends on the integer 
number of excess electrons $n = \pm1, \pm2 , ...$ on the island and on the
continuously varied voltages. 
Elementary electrostatics yields the ``charging energy''
\begin{equation}
	E_{\rm ch}(n,Q_{\rm G}) = \frac{(ne-Q_{\rm G})^2}{2C} \; .
\label{Ech}
\end{equation}
Here $C = C_{\rm L} + C_{\rm R} + C_{\rm G}$ 
is the total capacitance of the island.
The effect of the voltage sources is contained in the gate-charge 
$Q_{\rm G} = C_{\rm G} V_{\rm G} + C_{\rm L} V_{\rm L} 
+ C_{\rm R} V_{\rm R}$. Similar expressions hold for
the single-electron box.

In a tunneling process which changes the number of excess electrons
in the island from $n$ to $n+1$, the charging energy changes.
Tunneling from the left junction to the island is possible at low temperatures
only if the energy in the left lead, $eV_{\rm L}$, is high
enough to compensate for the increase in charging energy
$eV_{\rm L} > E_{\rm ch}(n+1,Q_{\rm G})- E_{\rm ch}(n,Q_{\rm G})$.
Similarly, tunneling from the island (transition from $n+1$ to $n$)
to the right lead is possible at low temperatures only if
$E_{\rm ch}(n+1,Q_{\rm G})- E_{\rm ch}(n,Q_{\rm G}) > eV_{\rm R}$.
Both conditions have to be satisfied simultaneously for a
current to flow through the transistor.
If this is not the case, the current is exponentially suppressed
realizing the so-called 
{\em Coulomb blockade}. Varying the gate voltage
produces Coulomb oscillations, i.e. an $e$-periodic dependence
of the conductance on $Q_{\rm G}$.
To further understand the characteristic of a SET transistor we need to 
determine the tunneling rates associated to electron tunneling.
An electron tunnels from one of the states $k$ in the
left lead into one of the available states  $q$ in the island,
thereby changing the electron number from $n$ to $n+1$, with rate
$\gamma_{\rm LI}$.
Such a rate, which is calculated by means of the Fermi golden rule, can be 
expressed in a transparent way as
\begin{equation}
	\gamma_{\rm LI}(n,Q_G) = \frac{1}{e} 
	I_{\rm qp}\left(\delta E_{\rm ch}/e\right)
	\frac{1}{\exp(\delta E_{\rm ch}/k_{\rm B} T) - 1} \; .
\label{ratesc}
\end{equation}
The function $I_{\rm qp}(V)$ is the  quasi-particle tunneling
characteristic (see e.g. Ref. \cite{tinkham96}), which is suppressed at voltages 
below the superconducting gap(s) and $\delta E_{\rm ch}$ is the charging energy difference
between the states with $n$ and $n+1$ electrons in the island. Charging effects reduce 
the quasi-particle tunneling further.  At low temperatures, $k_{\rm B} T \ll |\delta E_{\rm ch}|$,
a tunneling  process which  would increase the charging energy is suppressed.
This phenomenon is called {\em Coulomb Blockade} of electron tunneling.
At zero temperature the rate is nonzero only if the gain in charging energy 
compensates the energy needed to create the excitations
$\delta E_{\rm ch} + 2 \Delta  \le 0$.

The rates describe the stochastic time evolution of the charge of 
the junction system. For the theoretical analysis a master equation approach 
can be used. Several examples of the current-voltage characteristics of normal 
metal can be found in  Refs. \cite{averin,grabert}, an important
characteristic is the $e$-periodic dependence of the current and
conductance on the applied gate charge. 

As dictated by Eq. (\ref{ratesc}), the rate for quasi-particle tunneling 
are exponentially suppressed below the superconducting gap.
In this case higher-order processes involving multi-electron
tunneling can play a role. Indeed, for NS interfaces, 
there still exists a 2-electron tunneling process, in many respects similar 
to the Andreev reflection discussed in the previous section and for 
this reason denoted as Andreev tunneling.
The rate for Andreev tunneling taking into account charging effects was discussed 
in Ref. \cite{Hekking93a}.  Andreev tunneling is a second-order coherent process. In the first
part of the transition one electron is transferred from an initial state, e.g.
$k\uparrow$ of the normal lead, into an intermediate excited state
$q\uparrow$ of the superconducting island. In the second part of the coherent
transition, an electron tunnels from $k'\downarrow$
into the partner state $-q\downarrow$ of the
first electron, such that both form a Cooper pair.
The final state contains two excitations in the normal lead and an extra Cooper 
pair in the superconducting island, and the rates reads:
\begin{equation}
	\gamma^A_{\rm LI}(n,Q_{\rm G}) = \frac{G^{\rm A}}{4e^2}
	\frac{\delta E_{\rm ch,2}} {\exp (\delta E_{\rm ch,2}/k_{\rm B} T)-1} \; .
\label{rateA}
\end{equation}
Note that  the functional dependence of this rate coincides 
with that for single-electron tunneling in a normal junction,
Eq. (\ref{ratesc}), with a linear $I-V$ characteristics. Hence Andreev reflection is 
subject to Coulomb blockade like normal-state single-electron tunneling~\cite{Guinea}
with the exception that the charge transferred in an Andreev reflection is $2e$,
and the charging energy changes accordingly. 
The effective Andreev conductance is of second-order in the tunneling conductance 
and, as shown in Ref. \cite{Hekking93b}, is sensitive to spatial correlations in the normal
metal, which can be expressed by the Cooperon propagator.

\begin{figure}
\vspace{0.5cm}
\centerline{\psfig{figure=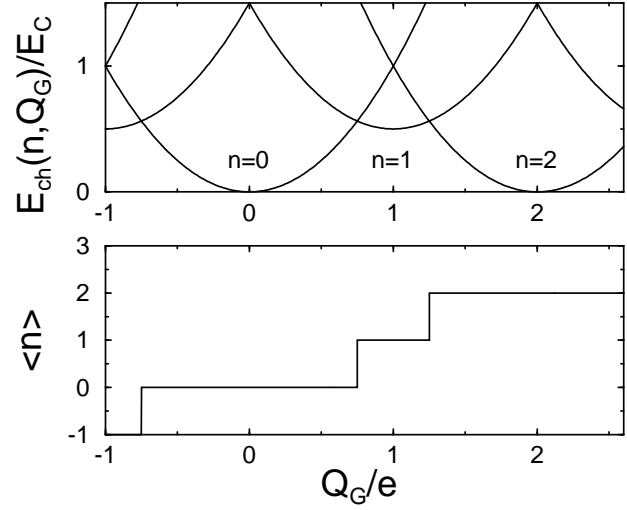,width=8cm}}
\caption{The charging energy of a superconducting
single-electron box as a function of the gate voltage
shows a difference between even and odd
numbers $n$ of electron charges on the island. Accordingly the
average island charge $\langle n \rangle$ is found in a broader range of
gate voltages in the even state than in the odd state.}
\label{figsuperparab}
\end{figure}

In a normal-metal electron box (see Fig. \ref{transistor}b), by sweeping the 
applied gate voltage the the number of electrons on the island in unit of $e$ and, as a 
consequence, the voltage of the island shows a periodic saw-tooth behavior~\cite{lafarge91}. 
The periodicity in the gate charge $Q_{\rm G}$ is $e$. If the island is
superconducting and the gap $\Delta$ smaller than the charging energy $E_{\rm C}$, 
the charge and the voltage show  at low temperatures a characteristic long-short cyclic, 
$2e$-periodic dependence on $Q_{\rm G}$. This
effect arises because single-electron tunneling from the ground state, where
all electrons near the Fermi surface of the superconducting island are
paired, leads to a state where one extra electron -- the ``odd'' one
-- is in an excited state~\cite{Averin92}. In a small island, 
as long as charging effects prevent further tunneling, the odd electron does not find another 
excitation for recombination. Hence the energy of this state stays
above that of the equivalent normal system by the gap energy. Only at larger
gate voltages can another electron enter the island, and the system can relax 
to the ground state. This scenario repeats with periodicity $2e$ in $Q_{\rm G}$, 
as displayed in Fig.~\ref{figsuperparab}.

Parity effects appear at temperatures lower than a crossover temperature 
$T_{\rm eff}$ which, in the experiments, is typically of the order of $10-30 \%$
of $T_{c}$. At low temperatures the even-odd asymmetry has been observed
in the single-electron box~\cite{Lafarge} as well as in the $I$-$V$ characteristics 
of superconducting SET transistor~\cite{Tuominen,Eiles,Hergenrother}.
However, at higher temperatures, above a cross-over value $T_{\rm cr} \ll \Delta$,
the $e$-periodic behavior typical for normal-metal systems is recovered. 

\subsection{NSN Transistor}

The analysis sketched above can be used to analyze the $I$-$V$ characteristics 
of  SET transistors with a superconducting island, the so-called NSN transistor. 
The interplay of 
parity effects and Andreev tunneling makes the current-voltage characteristics quite 
rich. If the energy gap is smaller than the charging energy the important processes are 
single-electron tunneling processes in the left and right junction, causing transitions
between even and odd states. In the opposite limit of a superconducting 
gap bigger than the charging energy,
the odd states have a large energy and Andreev tunneling, with rate given by Eq. (\ref{rateA}), 
becomes important~\cite{Hekking93a}.

At low temperatures, and superconducting gap larger than the charging energy, a set of 
parabolic current peaks is found centered around the degeneracy points 
$Q_{\rm G} = \pm e, \pm 3e, \ldots $ \cite{Hekking93a}
\begin{equation}
	I^{\rm A}(\delta Q_{\rm G}, V) = G^{\rm A}
	\left(V - 4\frac{\delta Q_{\rm G}^2}{VC^2}\right) 
	\Theta \left(V - 4\frac{\delta Q_{\rm G}^2}{VC^2}\right)\ \ .
\label{iandreev}
\end{equation}
Here $\delta Q_{\rm G} = Q_{\rm G}-e$ for $Q_{\rm G}$ close to e,
and similarly near the other degeneracy points.
At larger transport voltages, single-electron tunneling sets in,
even if $\Delta > E_{\rm C}$, and Andreev reflection
gets ``poisoned'' \cite{Hekking93a}. This occurs for
\begin{equation}
	  V \ge V_{\rm poison}
	= \frac{2}{e} \left( E_{\rm C} 
	- \frac{e Q_{\rm G}}{C} + \Delta \right) \; .
\end{equation}

\begin{figure}
\vspace{-2cm}
\centerline{\psfig{figure=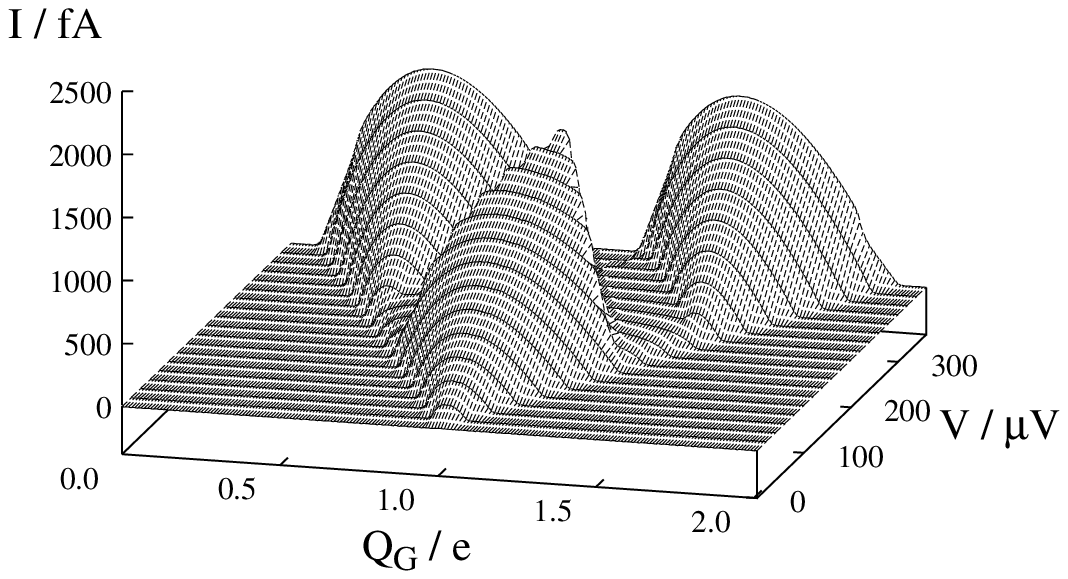,width=13cm}}
\vspace{-1cm}
\caption[]{ The current $I(Q_{\rm G},V)$ through a 
NSN transistor  with $\Delta > E_{\rm C}$. The
parameters  correspond to those of the experiments 
\cite{Hergenrother}, $E_{\rm C}=100\mu e$V, $\Delta=245 \mu e$V,
$R_{\rm tl/R} = 43k\Omega, 1/G^{\rm A} \approx 1.2 (2.4) 10^8
\Omega$ for the left (right) junction.
From Ref.~\cite{SSZ}.}
\label{figssz3}
\end{figure}

Fig.~\ref{figssz3} shows the current-voltage
characteristic of a NSN transistor with $\Delta > E_{\rm C}$~\cite{SSZ}.
At small transport voltages the 2e-periodic peaks due to Andreev reflection
dominate; they get poisoned above a threshold voltage. The peaks at larger
transport voltages arise from a combination of single-electron tunneling
and Andreev reflection.
The shape and size of the even-even Andreev
peaks and some of the single-electron tunneling features
at higher transport voltages agree well with those observed in the 
experiments of Hergenrother et al. \cite{Hergenrother}.

\section{Conclusions}
\label{conc}
In this review we tried to present an overall description of mesoscopic hybrid structures. 
The field is so vast that the choice of the topics was surely biased by our personal taste
and by our field of investigations.
To summarize, we have started this review by illustrating the transport properties of
hybrid systems formed by
putting into contact superconductors with normal metals and ferromagnets.
In Section II we have introduced the Bogoliubov-de Gennes equation in order
to discuss transport properties in terms of Andreev processes.
We have then provided a derivation of the Bogoliubov-de Gennes equation starting from
the BCS theory and given the solutions of such an equation for two different systems, 
namely a homogeneous superconductor and a ferromagnet/
superconductor junction.
After that we have introduced the two most important theoretical approaches to the
study of transport properties in hybrid systems.
We chose to give a more detailed treatment of the scattering theory, deriving
general conductance formul\ae~in terms of scattering matrices and giving explicit
expressions for the case of two-probe systems.
The quasi-classical Green's function approach has also been discussed, although in
less details, and a review of the most important results has been provided.
In Section II we have considered the heat transport in SN and SF structures to discuss the very efficient electronic cooling properties of hybrid systems.
Section III has been devoted to the Coulomb blockade, which is present in small metallic grains where charging effects are important, in hybrid systems.
We have discussed parity effects in single-electron transistors containing a 
superconducting island and their consequences on the quasi-particle tunneling
rates.
I-V characteristics has been furthermore illustrated.

There are several important topics which were left out,
the interested reader may find additional references in the reviews quoted in the
Introduction.

It is quite hard to draw conclusions and even harder to advance ideas about possible 
future directions of the field. Nevertheless we take this risk and briefly outline what 
we believe (remaining conservative) will be the future development of the field.
A direction which is not fully studied up to now is the role of superconductivity in molecular electronics. Transport through molecules using superconducting electrodes may give additional insight due to the fact that one can perform both single-particle and Cooper pair spectroscopy.

\acknowledgments 

We would like to acknowledge, on the topics covered in this review, a very fruitful 
collaboration with C. Bruder, G. Falci, F.W.J. Hekking, C.J. Lambert, G.M. Palma, R. Raimondi,
G. Sch\"on and J. Siewert. 
This work was supported by the EU (IST-SQUBIT, HPRN-CT-2002-00144)

\newpage

\newpage

\newpage

\newpage

\newpage

\begin{thebibliography}{299}
\bibitem{SNS} We wish to remark that the Bogoliubov-de Gennes equation was applied to
SNS Josephson junctions much earlier \cite{kulik70,bardeen72}.
However these works cannot be considered to fall within the scattering theory of
transport we are concerned with in the present section.
The reason is that they deal with equilibrium properties, like the Josephson current,
calculated through a thermodynamic approach.
Unlike in the Landauer-B\"uttiker scattering theory,
no scattering amplitudes describing a quasi-particle flow were defined.
\bibitem{ARED} See however a recently appeard paper \cite{samuelsson03} where the
Andreev reflection eigenvalue density was derived.
\bibitem{Nambu-Keldysh} 
	The Usadel equation can be written in terms of Green's function in the
	Nambu-Keldysh space (denoted by a check) as	
	$D\partial_{\mathbf{r}}(\check{G}\partial_{\mathbf{r}}\check{G})-i[\check{H},\check{G}]=0$,
	where $\check{H}=E\hat{\sigma}_3+i\hat{\sigma}_1\Re\Delta+i\hat{\sigma}_2\Im\Delta$.
\end{thebibliography}
\end{document}